\documentclass[11pt]{article} 

\usepackage[top=1in,bottom=1in,left=1in,right=1in]{geometry} 
\RequirePackage{amsthm,amsmath,amsfonts,amssymb}
\RequirePackage[numbers]{natbib}
\RequirePackage[colorlinks,allcolors=blue,urlcolor=blue]{hyperref}
\RequirePackage{graphicx}

\usepackage{color, amsmath, amssymb, amsbsy, amsthm, graphicx, bbm, amsfonts, bm, dsfont, courier, subfig}
\usepackage[toc,page]{appendix}
\usepackage{booktabs}
\usepackage[flushleft]{threeparttable}
\usepackage{latexsym}
\usepackage[dvipsnames]{xcolor}
\usepackage{comment}
\usepackage{wrapfig}
\usepackage{graphicx}
\usepackage{graphics}
\usepackage{algorithm}
\usepackage{algorithmic}
\usepackage{setspace}
\usepackage{adjustbox}

\usepackage{soul}
\usepackage{threeparttable,booktabs}
\usepackage{etoolbox}
\appto\TPTnoteSettings{\footnotesize}
\newtheoremstyle{theoremstyle}
{\topsep} 
{\topsep} 
{\itshape} 
{} 
{} 
{} 
{.5em} 
{\color{black}\ifthenelse{\equal{#3}{}}{{\bfseries #1 #2}}{{\bfseries #1 #2 (#3)}}}
\newtheoremstyle{theoremstylealt}
{\topsep} 
{\topsep} 
{\itshape} 
{} 
{} 
{} 
{.5em} 
{\color{black}\ifthenelse{\equal{#3}{}}{{\bfseries #1 #2$^\prime$}}{{\bfseries #1 #2$^\prime$ (#3)}}}
\newtheoremstyle{examplestyle}
{\topsep} 
{\topsep} 
{} 
{} 
{} 
{} 
{.5em} 
{\color{black}\ifthenelse{\equal{#3}{}}{{\bfseries #1 #2}}{{\bfseries #1 #2 (#3)}}}
\theoremstyle{theoremstyle}\newtheorem{thm}{Theorem}
\theoremstyle{theoremstylealt}
\theoremstyle{theoremstyle}     
\theoremstyle{theoremstyle}\newtheorem{lem}{Lemma}  
\theoremstyle{theoremstyle}\newtheorem{coro}{Corollary}        
\theoremstyle{theoremstyle}
\theoremstyle{theoremstyle}\newtheorem{assumption}{Assumption}
\theoremstyle{theoremstylealt}
\theoremstyle{theoremstyle}

\theoremstyle{theoremstyle}

\theoremstyle{theoremstyle}

\theoremstyle{examplestyle}\newtheorem{example}{Example}
\theoremstyle{examplestyle}
\theoremstyle{examplestyle}

\usepackage{xr}
\def\independenT#1#2{\mathrel{\rlap{$#1#2$}\mkern2mu{#1#2}}}

\renewcommand{\epsilon}{\varepsilon}

\def \hat{\widehat}

\newcommand\independent{\protect\mathpalette{\protect\independenT}{\perp}}
\def\independenT#1#2{\mathrel{\rlap{$#1#2$}\mkern2mu{#1#2}}}

\usepackage{titletoc}
\usepackage{enumitem}

\usepackage{array}
\newcolumntype{H}{>{\setbox0=\hbox\bgroup}c<{\egroup}@{}}

\usepackage{minitoc}

\def \hm{\hat{\texttt{M}}}
\def \m{\texttt{M}}
\def \hat{\widehat}
\def \tilde{\widetilde}

\def\x{\mbox{x}}
\def\z{\mbox{z}}
\def\y{\mbox{y}}
\def\w{\mbox{w}}
\def\t{\mbox{t}}
\def\s{\mbox{s}}
\def\D{\mbox{D}}
\def\transpose{^\intercal}

\def \hm{\hat{\texttt{M}}}
\def \m{\texttt{M}}
\def \hat{\widehat}
\def \tilde{\widetilde}

\def\x{\mbox{x}}
\def\z{\mbox{z}}
\def\y{\mbox{y}}
\def\D{\mbox{D}}
\def\s{\mbox{s}}
\def\w{\mbox{w}}
\def\t{\mbox{t}}
\def\hf{\hat{f}}
\def\transpose{^\intercal}

\begin{document}

\doparttoc 
\faketableofcontents 

\title{Assessing the Most Vulnerable Subgroup to Type II Diabetes Associated with Statin Usage: Evidence from Electronic Health Record Data
}
\author{Xinzhou Guo \thanks{Department of Mathematics, Hong Kong University of Science and Technology} \thanks{These authors contributed equally to this work and are alphabetically ordered.}  \and
Waverly Wei\thanks{Division of Biostatistics, University of California, Berkeley.} \footnotemark[2] \and
Molei Liu \thanks{Department of Biostatistics, Columbia Mailman School of Public Health} \and
Tianxi Cai\thanks{Department of Biostatistics, Harvard T.H. Chan School of Public Health}\and Chong Wu  \thanks{Department of Biostatistics, The University of Texas MD Anderson Cancer Center.} \and Jingshen Wang \footnotemark[3] \thanks{Correspondence: jingshenwang@berkeley.edu}
} 

\date{}

\maketitle

		\begin{abstract}
There have been increased concerns that the use of statins, one of the most commonly prescribed drugs for treating coronary artery disease, is potentially associated with the increased risk of new-onset type II diabetes (T2D). Nevertheless, to date, there is no robust evidence supporting as to whether and what kind of populations are indeed vulnerable for developing T2D after taking statins. In this case study, leveraging the biobank and electronic health record data in the Partner Health System, we introduce a new data analysis pipeline and a novel statistical methodology that address existing limitations by (i) designing a rigorous causal framework that systematically examines the causal effects of statin usage on T2D risk in observational data, (ii) uncovering which patient subgroup is most vulnerable for developing T2D after taking statins, and (iii) assessing the replicability and statistical significance of the most vulnerable subgroup via a bootstrap calibration procedure. Our proposed approach delivers asymptotically sharp confidence intervals and debiased estimate for the treatment effect of the most vulnerable subgroup in the presence of high-dimensional covariates. With our proposed approach, we find that females with high T2D genetic risk are at the highest risk of developing T2D due to statin usage.
 	 \\ \bigskip 
		
		\end{abstract}
		
		\vskip 2cm
		\begin{center}\bfseries
		\end{center}

\clearpage

\doublespacing

\section{Introduction}\label{Section-introduction}

\subsection{Motivation and objectives}

Coronary artery disease (CAD), a disease affecting the function of heart, is the leading cause of deaths worldwide \citep{skourtis2020nanostructured}. Over the past decades, efforts have been made in developing effective and safe drugs in preventing and treating CAD \citep{povsic2017navigating}. Among those novel agents, statins are perhaps the most commonly prescribed drugs due to their clear benefits in reducing the level of low--density lipoprotein (LDL) and subsequently lowering CAD risks through 3-hydroxy-3-methylglutaryl-coenzyme A reductase (HMGCR) inhibition \citep{nissen2005statin}.
Despite their clear benefits in reducing CAD risks, the use of statins is potentially associated with the increased risk of new-onset type II diabetes (T2D) \citep{waters2013cardiovascular,macedo2014statins,mansi2015statins}. 

Although many studies have been conducted to investigate the potential side effects of statins in developing T2D, to date, there is still no robust evidence as to whether and on what kind of populations statin usage increases the risk of T2D. Take some frequently cited studies as examples. \cite{rajpathak2009statin} find through meta-analysis that there is a small increase in T2D risk \footnote{Relative risk (RR): 1.13, 95\% CI [1.03, 1.23]} associated with the use of statins, but this association is no longer significant after including the results from the WOSCOPS trial \citep{packard1998influence}--the first study investigating the association between T2D and statins. Recent studies also suggest that the effect of statins on T2D risk might be heterogeneous across different sub-populations and be more pronounced in certain subgroups defined by sex and baseline T2D genetic risk \citep{mora2010statins, goodarzi2013relationship}. Nevertheless, existing studies may not lead to trustworthy findings in subgroups, because their statistical analyses either are conducted under randomized controlled trials (RCTs) with limited sample sizes whose results might not be generalizable beyond study population\citep{mora2006justification,mora2010statins}, or do not adjust for multiple comparisons issue when several candidate subgroups are under consideration \citep{waters2013cardiovascular}. 

In this case study, leveraging Partner Health System (PHS) biobank and electronic health record (EHR) data, we conduct subgroup analysis and assess the most vulnerable subgroup to T2D associated with statin usage from a novel biological perspective. We focus on the most vulnerable subgroup not only because pursuing the subgroup with the largest (adverse) treatment effect is a conventional practice in clinical studies \citep{naggara2011problem,kubota2014phase}, but also because a comprehensive understanding of the most vulnerable subgroup to T2D risk associated with statin usage could support precise clinical decisions and effective actions concerning the prescription of statins \citep{mora2010statins,bornkamp2017model,guo2020inference}. Concretely, our study consists of three objectives: (i) designing a rigorous causal framework that systematically examines the causal effects of statin usage on T2D risk from observational data, (ii) uncovering which patient subgroup is most vulnerable for developing T2D after taking statins, and (iii) assessing the replicability and statistical significance of the most vulnerable subgroup via a bootstrap calibration procedure.

\subsection{Overview of research methods and findings}

To systematically examine the causal effects of statin usage on T2D risk from observational data, we propose a novel study design which not only circumvent common issues in RCTs but also alleviate the concern of unmeasured confounding bias.

On the one hand, while existing studies often investigate the adverse effects of stains on T2D risk in RCTs with limited sample sizes, our study design leverages the large PHS biobank with linked EHR data, providing robust evidence for assessing the adverse effect of statin usage. We extract and link genotype information together with diagnostics from consented subjects in the Partner Health System(PHS) biobank and EHR data, respectively. This leads to an EHR virtual cohort of $17,023$ subjects, a much larger cohort than those from usual clinical trials, and $337$ features; see Section \ref{Sec:data-description} for detailed descriptions. Compared to study cohorts enrolled in RCTs, our study cohort can be a more representative sample of the general population (see Table \ref{tab:demo} for the demographics of our study cohort). 

On the other hand, while causal conclusions derived from observational studies can be susceptible to unmeasured confounding and reverse causation bias, our study design adopts a randomly inherited single nucleotide polymorphism (SNP), rs12916-T, as a \textit{surrogate treatment variable} of statin usage to alleviate the concerns of those issues. rs12916-T is a reliable surrogate treatment variable because it resides in the \textit{HMGCR} gene encoding the drug target of statins, and has been recently used as an unbiased, unconfounded proxy for pharmacological action on the target of statins (i.e., HMG-CoA reductase inhibition) \citep{swerdlow2015hmg}. Furthermore, adopting a randomly inherited genetic variant at conception as a surrogate for statin usage in EHRs allows us to establish a clear temporal precedence between the treatment and T2D onset (which is a prerequisite to concluding causality, see \cite{holland1986statistics} for example). This avoids potential reverse causation issues. Lastly, because the surrogate treatment variable is naturally inherited, variables observed after birth are independent of rs12916-T and can at most be mediators that belong to a different causal pathway. We shall discuss the reasoning of using rs12916-T in more detail in Section \ref{subsec:study-design}. 

Leveraging the above study design, we further conduct subgroup analysis and assess if subjects in different subgroups carrying the rs12916-T allele (i.e. taking statins) have heterogeneous risks at developing T2D, and
to what extent the most vulnerable subgroup suffers from the side effect of statin usage. Inspired by study designs adopted in \cite{mora2010statins} and \cite{wang2013detecting}, we divide our study cohort into six pre-defined candidate subgroups based on sex and baseline T2D genetic risk profiles (measured by the number of risk alleles of variants rs35011184-A and rs1800961-T each individual carries), and aim to uncover the most vulnerable patient subgroup to statin usage and assess the statistical significance of the most vulnerable subgroup. While numerous methods have been proposed in subgroup analysis for identifying subgroups \citep{lipkovich2011subgroup,ma2017concave} and testing subgroup homogeneity \citep{shen2015inference,fan2017change}, in our case study, the primary objective is to make valid post-hoc inference on the most vulnerable subgroup.

The limitation of usual post-hoc inference on the most vulnerable subgroup is well recognized. Due to the winner's curse bias induced by multiple comparisons \citep{efron2011tweedie}, post-hoc inference often leads to false positive results \citep{thomas2017comparing}. Although several attempts have been made to address the winner's curse bias issue, existing procedures are either poorly grounded \citep{stallard2008estimation,rosenkranz2016exploratory} or tend to be conservative \citep{hall2010bootstrap, fuentes2018confidence}, as the latter is typically built on simultaneous inference aiming to control the family-wise error rate for all candidate subgroups. These conservative simultaneous inference procedures are usually undesirable in subgroup analysis, because they yield false negative discoveries and have inadequate power to confirm the most vulnerable subgroup \citep{magnusson2013group, burke2015three}. In our context, because subgroup analysis needs to be conducted in observational studies \citep{yang2020propensity,lu2018estimating}, the problem becomes even more challenging as we need to take possibly high-dimensional confounders into account in assessing subgroup treatment effects.
To address the above-mentioned issues, we provide new post-hoc inferential tools to help assess the efficacy of the most vulnerable subgroups from observational studies without having to resort to simultaneous inference methods, which are often too conservative to start with. 

By applying the proposed method to our study cohort, we find that, although the overall adverse effect of statin usage on developing T2D is not significant, the female subgroup with high-genetic baseline T2D risk (more than two T2D risk alleles) is identified as the most vulnerable subgroup, meaning that with statin usage, this subgroup has the highest risk of developing T2D. The statistical significance of such a finding is also confirmed by the proposed bootstrap calibration method. In sum, our case study not only provides new evidence supporting the adverse effect of statin usage from a biological perspective, but also suggests that more caution should be taken when statins are prescribed, especially for females who are already at higher risk of developing T2D. The specific actions may include preventive treatments for diabetes and recommendations on lifestyle changes. 


\section{Data description and model setup}

\subsection{Study design}\label{subsec:study-design}
In our case study, since patients' statin use information is not available, we adopt the genetic variant (rs12916-T) as the surrogate treatment variable of statin usage. When the treatment indicator variable $\t = 1$, this means that ``the subject carries the variant rs12916-T." When $\t = 0$, this means that ``the subject does not carry the variant rs12916-T." We adopt this genetic variant as a surrogate treatment variable not only because statin usage information is not available in our EHR data, but also because carrying rs12916-T is a proxy for statin usage. The reason for adopting this proxy is due to the fact that rs12916-T, which resides in the \textit{HMGCR} gene encoding the drug target of statins, has been recently adopted as an unbiased, unconfounded proxy for pharmacological action on the target of statins \citep{swerdlow2015hmg}. In other words, rs12916-T allele and statins are functionally equivalent in that they both lower LDL cholesterol level through HMG-CoA reductase inhibition. Concretely, \cite{wurtz2016metabolomic} shows that the metabolic changes (e.g. decreased LDL cholesterol level) associated with statin usage have been found to resemble the association between rs12916-T and the metabolic changes with $R^2 = 0.94$. Therefore, we believe that rs12916-T is a credible surrogate treatment variable of statin usage. 

Furthermore, besides the absence of statin usage information in our EHR data, we shall argue that adopting rs12916-T as the surrogate treatment variable of statin usage invests our framework with two other key benefits.

\begin{figure}  
    \centering
 \includegraphics[width=0.5\textwidth]{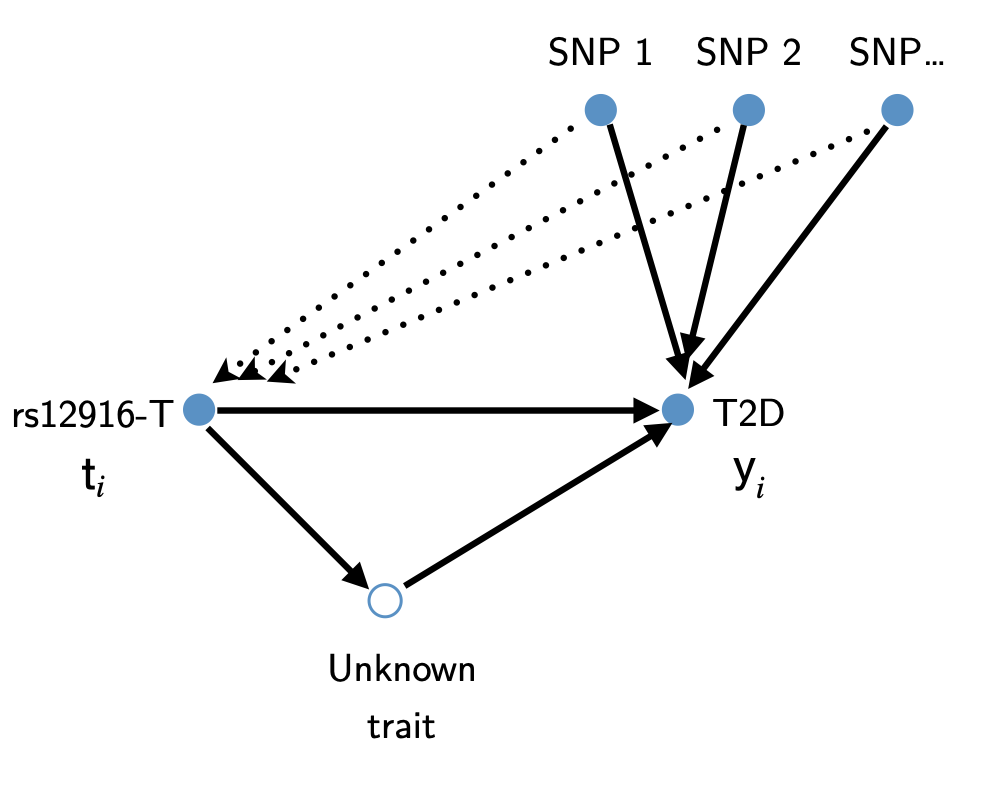}\caption{ The causal diagram under our study design.}
    \label{fig:causal-main}
\end{figure}

{First}, because our study design ensures the temporal precedence between inheriting the variant rs12916-T (surrogate of statin use) and T2D onset, this lifts the concern of reverse causation in conducting causal inference from observational data. In particular, when studying the causal effect of statin use on T2D risk from observational data, one normally assumes that statin use causes the change in T2D risk \citep{pan2020ldl}. This suggests that only data collected from subjects who have recorded statin use status before T2D onsets can be used for credible casual analyses. Unfortunately, in our EHR data, it is impossible to establish the temporal precedence between statin use and T2D onset. Using a genetic variant (rs12916-T) as the surrogate treatment variable circumvents the above-mentioned issue. Because genetic variants are randomly inherited at conception, our study design guarantees that the cause (carrying rs12916-T [proxy for pharmacological action of statin use] or not) must occur before T2D onset. This clear temporal precedence makes the established causal relationships more plausible (see Figure \ref{fig:causal-main}). 
 
{Second}, adopting rs12916-T as a surrogate treatment variable attenuates the concern of unmeasured confounding biases in observational data. Given rs12916-T is naturally inherited, variables observed after birth can at most be mediators that belong to a different causal pathway. Since we work with the causal pathway between the treatment and the outcome, i.e., $\t_i\rightarrow\y_i$, conditional on such (unmeasured) mediators is not necessary. Thus, the unmeasured confounding bias issue is alleviated under our study design. To further robustify our causal conclusion and improve statistical estimation efficiency of the underlying causal effect, we include additional potential confounders that are obtained before birth, such as genetic variants associated with T2D and potentially associated with rs12916-T. We provide more discussions on causal pathways in Supplementary Materials Section H.

Our outcome of interest is T2D status. We defer the technical details on defining T2D status from EHR data to Section \ref{Sec:data-description}. Given that the existing literature \citep{mora2010statins,waters2013cardiovascular} has suggested that the treatment effect of statin usage on T2D risk could be heterogeneous, we conduct subgroup analysis to investigate the causal effect of T2D associated with statin usage in subgroups and the most vulnerable one in particular. 
Inspired by the study designs in \cite{mora2010statins} and \cite{waters2013cardiovascular} in which patient population is divided based on sex in the former and the number of T2D baseline risk factors in the latter, we divide our PHS study cohorts into six pre-defined candidate subgroups based on sex and baseline T2D genetic risk profiles. The baseline T2D genetic risk is measured by the number of copies of T2D risk allele of variants rs35011184-A and rs1800961-T each subject has. The more T2D risk alleles the subject has, the higher baseline genetic T2D risk the subject bears \citep{lango2008assessing}. We define ``low-risk" as the total number of alleles $=0$, ``mid-risk" as  the total number of alleles $=1$, and ``high-risk" as the total number of alleles $\geq 2$. The six subgroups are thereby divided as (1) high-risk female; (2) mid-risk female; (3) low-risk female; (4) high-risk male; (5) mid-risk male, and (6) low-risk male.

Here, we consider pre-defined subgroups instead of post-hoc identified subgroups because pre-defined subgroups usually have clearer interpretability and could avoid the bias induced by data-adaptive subgroup identification procedure, while post-hoc identified subgroups are often adopted when there is no prior information on the segregation of study population \citep{lipkovich2011subgroup, ma2017concave}. In our setting, because previous studies \citep{mora2010statins,waters2013cardiovascular} suggest that T2D risk might have differential effects across sex and baseline T2D genetic profiles, pre-defined subgroups are more suitable for the present case study. In Supplementary Materials Section G, we compare the pre-defined subgroups and post-hoc identified subgroups based on our EHR data, and we find that the post-hoc identified subgroups resemble the pre-defined subgroups adopted in our case study.

\subsection{Model setup}

We work with the following sparse logistic regression model:
\begin{equation}\label{eqn:setup}
    \text{logit}\left\{{\mathbb{P}}(\y=1\mid \z,\x)\right\} = \mbox{z}\transpose\beta + \x\transpose\gamma,\quad  \left\|\gamma\right\|_0 \ll  p.
\end{equation}
Here, $\y$ is the observed binary outcome representing the T2D status. $\z\in\mathbb{R}^{p_1\times n}$ includes variables representing
interactions between the treatment variable and all the six subgroup indicator variables.  $\x\in\mathbb{R}^{p_2\times n}$ contains $336$ covariates and an intercept (hence $p_2 = 337)$.
The $336$ covariates contain $5$ subgroup indicator variables (the sixth subgroup, low-risk male, indicator variable is dropped to avoid collinearity) and 331 potential confounders (including  race and age as baseline characteristics, and $329$ SNPs associated with T2D related factors accounting for potential confounding issues). Note that we do not include the treatment variable as a covariate because including it causes collinearity issues.
All observed covariates are obtained from Partner Health System biobank. 

Following the above setup, $\beta \in\mathbb{R}^{p_1}$ represents subgroup causal effects on the scale of log odds ratio (OR) (hence $p_1 = 6$). More concretely, under Model \eqref{eqn:setup}, $\beta = \big(\log \alpha_1, \ldots, \log \alpha_6\big)$ with $\log\alpha_j$ representing the log odds ratio of subgroup $j$, for $j=1, \ldots, 6$.  
Following the Neyman-Rubin causal model and our current study design, we provide rigorous causal identification results to justify why the model parametrization in \eqref{eqn:setup} enables us to estimate the heterogeneous causal effects in the pre-defined subgroups. This theoretical justification is provided in Supplementary Materials Section E. We further assume that $\gamma\in \mathbb{R}^{p_2}$ is a sparse vector with the support set $\texttt{M}_0$. The sparsity assumption not only provides a parsimonious explanation of the data but also carries our prior belief that not every genetic variant is predictive of the outcome as demonstrated in Section \ref{Sec:data-description}.

\subsection{Data description and exploration}\label{Sec:data-description}

Following our study design described in the previous section, we extract and link genotype information and diagnostics from consented subjects in the PHS biobank and EHR data respectively. Our data involve a much larger cohort than those from usual clinical trials, $n= 17,023$ subjects each with $p=337$ features; see Table \ref{tab:demo} for data summary. 



\begin{table}[h!]
    \centering
    \small{
\begin{tabular}{lc}
\hline
\hline
\\[-2.5ex] 
Variable           & Frequency (percent) \\
\\[-2.5ex] 
\hline
\\[-2.5ex] 
Sex                &                     \\
~~Female             & 7,592 (45)           \\
~~Male               & 9,431 (55)           \\
\\[-2.5ex] 
\hline
\\[-2.5ex] 
Age (years)        &                     \\
~~$<$40      & 2,333 (14)           \\
~~40--50   & 1,733 (10)           \\
~~50--60   & 2,824 (17)           \\
~~60--70   & 3,971 (23)           \\
~~70--80   & 4,042 (24)           \\
~~$\geq$80  & 2,120 (12)           \\
\\[-2.5ex] 
\hline
\\[-2.5ex] 
Race               &                     \\
~~European           & 15,048 (88)          \\
~~African American   & 1,004 (6)            \\
~~Other/Unknown      & 971 (6)             \\
\\[-2.5ex] 
\hline
\\[-2.5ex] 
Ethnicity          &                     \\
~~Hispanic or Latino & 698 (4)             \\
~~Other/Unknown      & 16,325 (96)      \\
\\[-2.5ex] 
\hline
Number of rs12916-T allele                &                     \\
~~2      &  6,245 (37)           \\
~~1              & 8,079 (47)           \\
~~0              &  2,699 (16)           \\
\\[-2.5ex] 
\hline
\\[-2.5ex] 
With T2D        &                     \\
~~Yes     & 2,565 (15)           \\
~~No  &  14,458 (85)           \\

\hline
\hline
\end{tabular}}
  \caption{\label{tab:demo}\small Demographics of 17,023 PHS subjects considered in our study.}
\end{table}

Recall that the covariates contain age, race, subgroup indicators, and genetic variants associated with T2D related factors (including LDL, high density lipoprotein and obesity). As for the definition of the outcome, since the diagnostic billing code for T2D has limited specificity in classifying the true T2D status, we define the T2D status based on a previously validated multimodal automated phenotyping (MAP) algorithm\citep{liao2019high}. The area under the ROC curve (AUC) of MAP's risk prediction score for classifying true T2D status is $0.99$, and the specificity and sensitivity of its classifier are $0.97$ and $0.92$ respectively. These suggest that the MAP classifier of T2D can be reliably used to define the T2D outcome. Among our study cohort, MAP classifies $2,565$ subjects having T2D. There is no missing data issue in our case study for two reasons. First, we leverage the large PHS biobank with linked EHR data, thus the genetic profiles and baseline covariates are non-missing. Second, we adopt surrogate outcomes, thus we do not encounter any missing outcomes.

To explore the association between statin usage and T2D risk, we report preliminary data exploration results from a ``full" logistic regression  model for $\y$ against the treatment $\mbox{t}$ and the covariates $\x$ in Table \ref{table:overall-statins}; i.e. $y\sim \mbox{t}+\x$.  
There, although a modest association is found between carrying the rs12916-T variant and T2D status in the overall study cohort, unlike the results in \cite{swerdlow2015hmg}, this association is not statistically significant. Moreover, our analysis reports only $16$ regression coefficients having $p$-values $<0.05$, suggesting that the logistic regression coefficient vector for the covariates is likely to be sparse. Because the overall treatment effect is marginal (estimated marginal treatment effect equals 0.04 with p-value 0.35), this motivates us to conduct subgroup analysis to further investigate the subgroup causal effect of statin usage on T2D risk.


\begin{table}[h!]
 \centering
 \begin{tabular}{cccccc}
    \hline\hline
     \\[-2.5ex] 
        Treatment effect & Est & SE & $Z$-value & $p$-value & \# of significant coefficients\\
        \\[-2.5ex] 
        \hline
    \\[-2.5ex] 
    Full logistic regression  & $0.04$ & $0.04$ &  $0.93$ &  $0.35$ & 16 \\
   \\[-2.5ex] 
      \hline
      \hline
    \end{tabular}
\caption{The estimated treatment effect (Est), standard error (SE), $Z$- and  (two-sided) $p$-values of statins' usage on the overall PHS study cohort. We fit the ``full" logistic regression model with $337$ features (age, race and genetic information). ``significant coefficients" are the estimated regression coefficients with $p$-values $<0.05$.
}
\label{table:overall-statins}
\end{table}

\subsection{Challenges in statistical inference}\label{sec: model-setup}

Because our goal is to assess the patient subgroup most vulnerable for developing T2D, our methodological development hence centers around delivering valid inference (accurate point estimate and valid confidence interval) on the effect size of maximal regression coefficient $\beta_{\max} = \underset{j \in\{1, \ldots, p_1\} }{\max}\beta_j$ in Model (1). 
We focus on $\beta_{\max}$ instead of $|\beta|_{\max}$ for the following reason. The logistic regression coefficient $\beta$ represents the log odds ratio, thus each regression coefficient is a number ranging from $-\infty$ to $\infty$. A larger $\beta$ indicates a higher T2D risk associated with statin usage. If we use $|\beta|$, a larger $|\beta|$ might no longer measure the adverse effect of statin usage on T2D risk, implying that $|\beta|_{\max}$ represents the adverse effect of either the most vulnerable subgroup or the least vulnerable subgroup to statin usage. Because $\beta_{\max}$ has clearer interpretation than $|\beta|_{\max}$, we focus on $\beta_{\max}$ instead of $|\beta|_{\max}$ in this manuscript.


In the presence of high-dimensional covariates as described in Section \ref{Sec:data-description}, finding an accurate point estimate and conducting inference on $\beta_{\max}$ can be a challenging task, due to the presence of regularization and winner's curse biases. The regularization bias occurs whenever penalization approaches are adopted to select a smaller working model to enhance the estimation efficiency of $\beta$ in the presence of sparsity \citep{wang2018debiased, hong2018overfitting}. The winner's curse bias occurs whenever we use a simple sample-analogue $\hat{\beta}_{\max}=\underset{j \in\{1, \ldots, p_1\} }{\max}\hat{\beta}_j $ to estimate the true maximum effect $\beta_{\max}$. A sample average estimate for $\beta_{\max}$ overestimates the parameter because, even if $\hat{\beta}$ follows normal distribution centering at $\beta$, $\hat{\beta}_{\max}$ will follow a skewed-normal distribution and will not center at $\beta_{\max}$ \citep{nadarajah2008exact,guo2020inference}. Such an overestimation phenomenon is well-recognized in post-hoc subgroup analysis \citep[see][for example]{zollner2007overcoming, cook2014lessons}. While several approaches have been proposed to address the regularization bias issue \citep{zhang2014confidence,li2020debiasing}, and provide valid inference on a single regression coefficient, these methods cannot account for the winner's curse bias. As for the winner's curse bias, existing methods are mostly made for low-dimensional data and are not directly applicable to observational data with high dimensional covariates in this case study \citep{bornkamp2017model,guo2020inference}. While \cite{guo2021sharp} proposes bootstrap-based approaches to simultaneously address the regularization and winner's curse bias issues in high dimensional linear models, we broaden its validity by providing a bootstrap procedure that is asymptotically valid for high dimensional logistic regression estimators with rigorous statistical guarantees. To our knowledge, debiasing procedures that simultaneously remove the regularization bias and winner's curse bias in high dimensional non-linear models have been lacking. Technical discussions on these bias issues are deferred to the Supplementary Materials Section A, and we demonstrate the winner's curse bias and regularization bias in estimating $\beta_{\max}$ within a simple simulated example.

\begin{figure}
        \centering\includegraphics[width=0.9\textwidth]{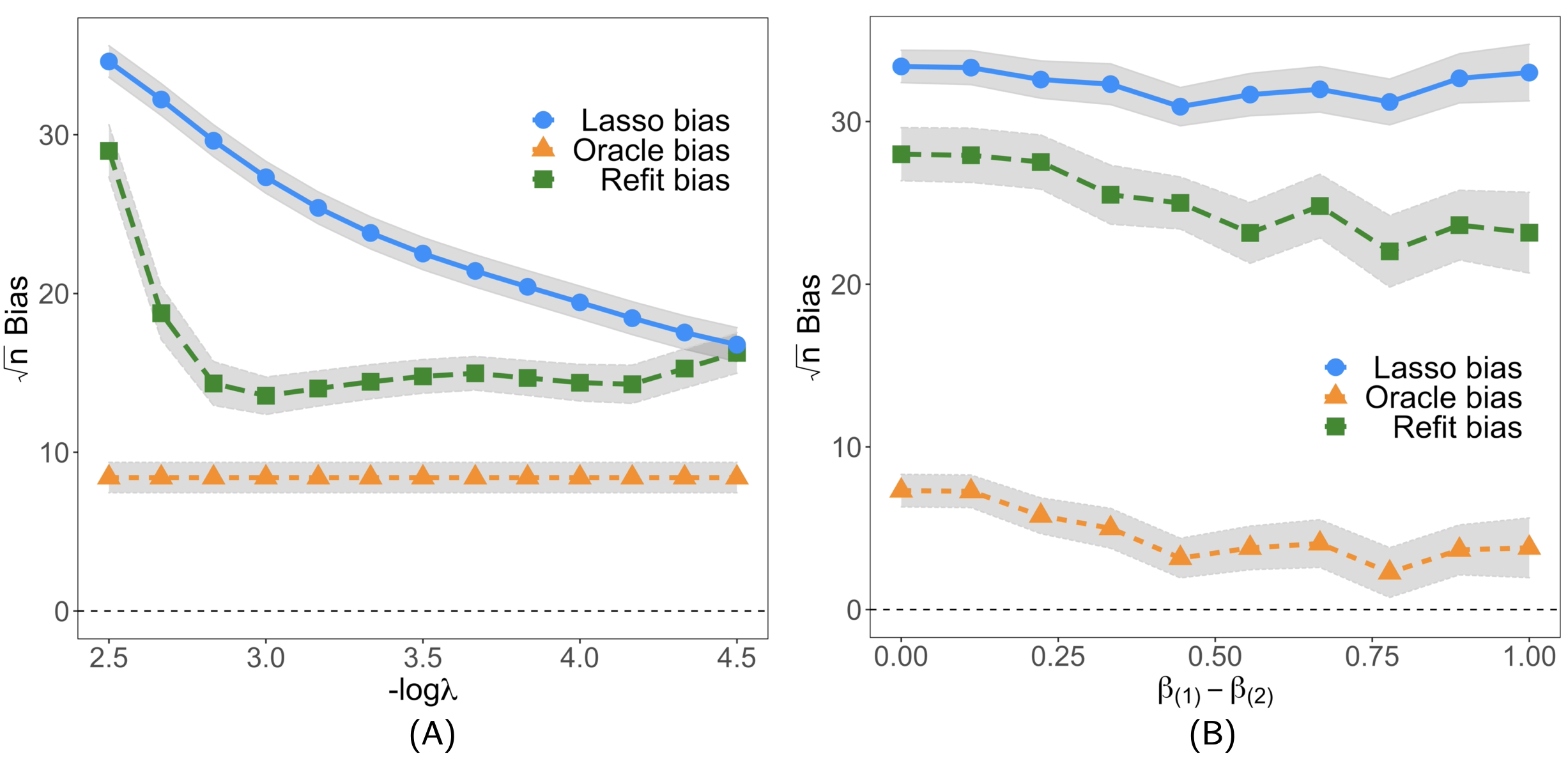}
        \caption{\small \label{fig:bias-example} Root-$n$ scaled bias for Example 1. The shaded areas are calculated based on Monte Carlo standard errors. 
        }
    \end{figure}

\begin{example}[Winner's curse bias and regularization bias in estimating $\beta_{\max}$]\label{example:bias}\normalfont  We use two widely adopted procedures to estimate $\beta$: (1) Lasso for generalized linear models (GLM) \citep{hastie2007glmlasso}, which estimates $\beta$ with $\big( \hat{\beta}_{\texttt{GLasso}}\transpose,\hat{\gamma}_{\texttt{GLasso}}\transpose\big)\transpose $ obtained from the $\ell_1$-penalized logistic regression program without any adjustments, and (2) Refitted GLM Lasso, which estimates $\beta$ by refitting the logistic regression model based on the covariates in the support set of $\big( \hat{\beta}_{\texttt{GLasso}}\transpose,\hat{\gamma}_{\texttt{GLasso}}\transpose\big)\transpose $. As a benchmark, we also report the performance of the oracle estimator $\big( \hat{\beta}_{\texttt{Oracle}}\transpose,\hat{\gamma}_{\texttt{Oracle}}\transpose\big)\transpose $ which pretends the true support set of $\gamma$ is known and is estimated by refitting the logistic regression model with the true support set. $\beta_{\max}$ is then estimated in a two-step procedure: One first obtains an estimate $\hat{\beta}$ and then estimates $\beta_{\max} $ by taking the maximum, that is $\max\{ \hat{\beta}_1, \ldots,\hat{\beta}_{p_1}\}$. To mimic the causal relationship in this case study, 
we generate Monte Carlo samples with $\t_{i} \sim \text{Bernoulli}(0.5)$ independent of the covariate $\mbox{w}_i \sim N( 0, \Sigma)$, where $\Sigma = (\Sigma_{jk})_{j,k=1}^{p-6}$ and $\Sigma_{jk} = 0.5^{|j-k|}$ for $i=1, \ldots, n$. We then generate $\x_{ij} = \mathds{1}(\mbox{w}_{ij}>0)$ for $1 \leq j \leq p-6$, and
$\z_{il} = t_ix_{il}$, $l=1,\ldots,6$. $\y_i$ is generated following Model \eqref{eqn:setup}. 
 We set the sample size $n=1,000$ and the dimension $p=200$, and set  $\gamma = (1, 1, 0, \ldots)\in\mathbb{R}^{p-6}$. For the first simulation, we set the coefficients  $\beta = (0.5, 0.5, 0, 0, 0, 0)\transpose$ and vary the value of tuning parameter $\lambda$ to illustrate how the winner's curse bias could invalidate the inference as the winner's curse bias is the most severe when the two largest $\beta$'s are equal \citep{guo2020inference,nadarajah2008exact}. The results are shown in Figure \ref{fig:bias-example} (A). For the second simulation, to demonstrate how the winner's curse bias changes with respect to the distance between the largest and the second largest components in $\beta$, i.e., $\beta_{(1)} - \beta_{(2)}$,
 we fix $\log\lambda = -2.5$ for illustration, set the coefficients  $\beta = (\beta_{\max}, 0.5, 0, 0, 0, 0)\transpose$, where $\beta_{\max}\in \{0.5,0.61,0.72, \ldots, 1.5\}$, and plot the $\sqrt{n}$-scaled bias with respect to various $\beta_{(1)}-\beta_{(2)}$ values, where $\beta_{(1)}$ is equivalent to $\beta_{\max}$. The results are presented in Figure \ref{fig:bias-example} (B). In Figure \ref{fig:bias-example}, we report the root-$n$ scaled bias based on $500$ Monte Carlo samples under the two settings respectively.
    \end{example}
 
From the results in Figure \ref{fig:bias-example}, we observe that all three estimators are biased. Although $\hat{\beta}_{\text{Oracle}}$ is a consistent estimator of $\beta$, its maximum is not centered around $\beta_{\max}$. Following some explicit evidence given in \cite{nadarajah2008exact}, $\hat{\beta}_{\max}$ is usually biased upward for estimating $\beta_{\max}$, we thus conjecture that the residual bias in the maximal of the oracle estimator $\hat{\beta}_{\text{Oracle}, \max}$ is caused by the winner's curse bias issue. We further observe that the magnitude of the winner's curse bias decreases as the distance between $\beta_{(1)}$ and $\beta_{(2)}$ increases (as seen in Figure \ref{fig:bias-example} (B)), suggesting that the winner's curse bias might not be a severe concern if $\beta_{(1)}$ and $\beta_{(2)}$ are far apart. As $\beta$ is unknown a priori, inference procedure  without adjusting for the winner's curse bias may not be valid in practice. On the top of the winner's curse bias issues, the GLM Lasso and the refitted estimators suffer from the regularization bias and hence are also not correctly centered around $\beta_{\max}$, unless in some special cases where the regularization bias and the winner's curse bias cancel out.

To simultaneously adjust for the winner's curse bias and the regularization bias without knowing the underlying true parameters, in what follows, we propose an inferential framework that produces a bias-reduced estimate as well as a valid confidence interval of $\beta_{\max}$.

\section{Methodology}\label{Sec:rsplit-boot}


 We start with describing an estimation strategy of $\beta$ that resolves the regularization bias induced by model selection and helps addressing our research objectives (ii) and (iii) discussed in Section 1.1. Regularization bias arises when the selected model is either over-fitted or under-fitted; see detailed discussion provided in the Supplementary Materials Section A. While the risk of under-fitting can be mitigated by aiming for a larger model for parameter estimation, we resolve the issue of over-fitting by sample splitting. Sample splitting divides a sample into two parts: The first part of the sample is used for model selection and the remaining part is used for estimation based on the selected model. When $\gamma$ is sparse and a larger model is selected on the first half of the sample, we expect refitted GLM estimator on the second part of the sample to be free of significant bias. Nevertheless, sample splitting provides debiased estimator of $\beta$ at a cost of increased variability, because only a part of the sample is used for estimation. To minimize this efficiency loss due to sample splitting, we consider the method of repeated sample splitting (R-Split) that averages different estimates of $\beta$ across different splits. Our strategy, in a spirit similar to bagging and ensemble algorithms in machine learning, helps to stabilize and improve the accuracy of the estimated $\beta$ in a subsample. 
 
 	\begin{description}
	\item[Step 1](Repeated sample splitting that accounts for the regularization bias) \textit{For $b\leftarrow 1$ to $B_1$:
	(1) Randomly split the sample} $\{ (\y_i, \x_i, \z_i) \}_{i=1}^n$ \textit{into two subsamples: a subsample $T_1$ of size $n_1$ and a subsample $T_2$ of size $n_2 = n-n_1$; (2) select a model $ \hm_b$ to predict $\y$  based on $T_1$; (3) refit the selected model with the data in $T_2$ to estimate $\beta_b$ and $\gamma_b$ via logistic regression}:
	    \begin{align*}
	        (\hat{\beta}_b\transpose  ,\hat{\gamma}_b\transpose)\transpose  =  \arg\min \left\{ \sum_{l\in T_2}\Big(\y_l\cdot( \z_l\transpose \beta + \x_{l, \hm_b}\transpose \gamma)-\log \big(1+\exp(\z_l\transpose \beta +\x_{l, \hm_b}\transpose \gamma)\big)\Big)\right\}.
	    \end{align*}
\textit{(4) obtain the R-Split estimate: $\tilde{\beta} = \frac{1}{B_1} \sum_{b=1}^{B_1} \hat{\beta}_b$.}
\end{description}

In this step, any reasonable model selection procedures may be used and the choice of model size is subjective, but the selected model needs to be large enough for the under-fitting bias to be negligible. In our simulation and case study, we use GLM Lasso for model selection \citep{friedman2017package} and choose the model size from cross-validation (see Supplementary Materials Section B for detailed description). 
The choice of splits $B_1$ needs to be sufficiently large so that the R-Split estimator $\tilde{\beta}$ has a tractable asymptotic distribution. Under appropriate regularity conditions, we show that $\tilde{\beta}$ converges to a normal distribution centered around $\beta$ at a root-$n$ rate (statistical justification is provided in the Supplementary Materials Section C.2).

As $\tilde{\beta}$ provides an accurate estimate of $\beta$, we use $\tilde{\beta}$ to address our research objectives (ii) and (iii). In particular, the subgroup with the largest coefficient, $\arg\max_{j\in[p_1]}\tilde{\beta}_j$, is most vulnerable for developing T2D after taking statins.
However, due to the winner's curse bias, simply relying on $\tilde{\beta}$ will not lead to valid inference on $\beta_{\max}$, and we need a second step to address objective (iii). Built upon an accurate estimate of $\beta$, we store an inverse Hessian matrix for the later bootstrap calibration to adjust for the winner's curse bias:
	$$\tilde{\Gamma}_n = \frac{1}{B_1}\sum_{b=1}^{B_1}    \mbox{I}_z \left( \frac{1}{n_1}\sum_{i\in T_{2,b} }  f_{i,b} \begin{pmatrix} \z_i \\
			\x_{i,\hm_b} \end{pmatrix} (\z_i\transpose , \x\transpose _{i,\hm_b})\transpose  \right)^{-1}\mbox{I}_{\hm_b},$$
where $f_{i,b} = \text{expit}'(\z_i\transpose \hat{\beta}_b + \x\transpose _{i,\hm_b}\hat{\gamma}_b)$. Its benefits will be apparent in the following step:

	


\begin{description}
	\item[Step 2] (Calibrated bootstrap that accounts for the winner's curse bias) \textit{For $b\leftarrow 1$ to $B_2$:
generate bootstrap replicate $\tilde{\beta}^*$ from:}
\begin{align}\label{eq:bootstrap-rsplit}
    \tilde{\beta}^* =  \tilde{\beta}+\tilde{\Gamma}_n \cdot \frac{1}{n} \sum_{i=1}^n \begin{pmatrix}
    \z_i\\
    \x_{i}
    \end{pmatrix} \nu^*_i,
    \end{align}
 \textit{where $\nu_i^*=u_i\hat{\nu}_i$ is the permuted GLasso residual}, $\hat{\nu}_i = \y_i - \text{expit}(\z_i\transpose\hat{\beta}_{\texttt{GLasso}}+\x_i\transpose\hat{\gamma}_{\texttt{GLasso}})$. \textit{Then recalibrate bootstrap statistics  via}
	    \begin{align*}
	        T_b^* =\underset{j\in [p_1]}{\max}(\tilde{\beta}^*_{j}+\tilde{c}_j(r))- \tilde{\beta}_{\max},\quad \tilde{c}_j(r) = (1-n^{r-0.5})(\tilde{\beta}_{\max} - \tilde{\beta}_j), \text{ where }r\in (0, 0.5). 
	    \end{align*}
\end{description}

In this step, rather than adopting the simple bootstrap statistics $\max_{j\in[p_1] } \tilde{\beta}^*_j - \tilde{\beta}_{\max}$ to make inference on $\beta_{\max}$, we make an adjustment to each coordinate of $\tilde{\beta}^*$ by the amount $\tilde{c}_j(r)$. This is because just as $\tilde{\beta}_{\max}$ is a biased estimator of $\beta_{\max}$, the simple bootstrap statistics $\max_{j\in[p_1] } \tilde{\beta}^*_j$ is also not centered at $\tilde{\beta}_{\max}$. The amount of adjustment $\tilde{c}_j(r)$ is large when $\tilde{\beta}_{j}$ is small, and is small when $\tilde{\beta}_j$ is large. By adding the correction term $\tilde{c}_j(r)$, under certain regularity conditions, the distributions of $\sqrt{n}(\tilde{\beta}^*_{\text{modified;max}} - \tilde{\beta}_{\max})$ and $\sqrt{n}(\tilde{\beta}_{\max} - \beta_{\max})$ are asymptotically equivalent, implying that our proposed method adjusts for the winner's curse bias and the regularization bias simultaneously, where $\tilde{\beta}^*_{\mathrm{modified};\max}= \underset{j\in [p_1]}{\max}(\tilde{\beta}^*_{j}+\tilde{c}_j(r))$. We relegate the theoretical details of this bootstrap calibration procedure in the Supplementary Materials Section C. Note that $r\in (0,0.5)$ is a positive tuning parameter (see Supplementary Materials Section B for its data adaptive choice).

At this point, we note that our procedure adopts wild bootstrap to construct bootstrapped statistics of the R-Split estimate $\tilde{\beta}$. The wild bootstrap procedure adopted here is not only computationally efficient in high dimensions, as the Hessian matrix remains unchanged across different bootstrap samples, but also provably consistent in our problem setup.  Furthermore, \cite{dezeure2017high} shows that wild bootstrap can be more versatile than other residual bootstrap methods because it correctly captures the asymptotic variance for various settings.
With the help of a valid bootstrap calibration procedure in replicating $\tilde{\beta}_{\max}$, we are now ready to propose our final step that constructs confidence intervals and debiased estimate for $\beta_{\max}$:

\begin{description}
	\item[Step 3] (Bias-reduced $\tilde{\beta}_{\max}$ and sharp confidence interval)  \textit{The level-$\alpha$ two-sided confidence interval for $\beta_{\max}$ is  $[\tilde{\beta}_{\max}-{Q}_{T^*_b}(\alpha/2), \tilde{\beta}_{\max}+{Q}_{T^*_b}(\alpha/2))$, and a bias-reduced estimate for $\beta_{\max}$ is  $\tilde{\beta}_{\max}-\frac{1}{B_2}\sum_{b=1}^{B_2}T_b^*$.}
	\end{description}

\section{Theoretical and empirical justification}\label{Section-theory}

In this section, we provide theoretical justifications of the proposed bootstrap-assisted R-Split estimator along with a simple power analysis, where we demonstrate that our approach not only has rigorous theoretical guarantee but also shows high statistical detection power. We then examine the performance of the proposed method through simulation studies. 
	
\subsection{Theoretical investigation and a power analysis}\label{subsec:theory-power}

The following theorem confirms that the asymptotic distribution of  $\sqrt{n}(\tilde{\beta}_{\text{modified;max}} - \tilde{\beta}_{\max})$ converges to $\sqrt{n}(\tilde{\beta}_{\max} -  \beta_{\max})$. This suggests that the proposed confidence interval constructed in Step 3 of Section \ref{Sec:rsplit-boot} is ``asymptotically sharp," meaning that it achieves the exact nominal level as the sample size goes to infinity. This distinguishes the proposed procedure from other conservative methods made for subgroup analysis \citep[for example]{hall2010bootstrap,fuentes2018confidence}. The proof of Theorem \ref{thm: theorem 1} is provided in the Supplementary Materials Section C.3.  To simplify presentation, we relegate regularity assumptions to the Supplementary Materials Section C.1.

\begin{thm}\label{thm: theorem 1}
	Under Assumptions 1-9 given in the Supplementary Materials Section C.1, when 
	$p_1$ is a fixed number, the modified bootstrap maximum treatment effect estimator, $\tilde{\beta}^*_{\mathrm{modified};\max}= \underset{j\in [p_1]}{\max}(\tilde{\beta}^*_{j}+\tilde{c}_j(r))$, satisfies:
	$$\sup_{c\in\mathbb{R}}|\mathbb{P}(\sqrt{n}(\tilde{\beta}_{\max}-\beta_{\max})\le c)-\mathbb{P}^*(\sqrt{n}(\tilde{\beta}^*_{\mathrm{modified};\max}-\tilde{\beta}_{\max})\le c)|=o_p(1).$$
\end{thm}

\begin{figure}
  \vspace{-0.3cm}
       \centering\includegraphics[width=0.45\textwidth]{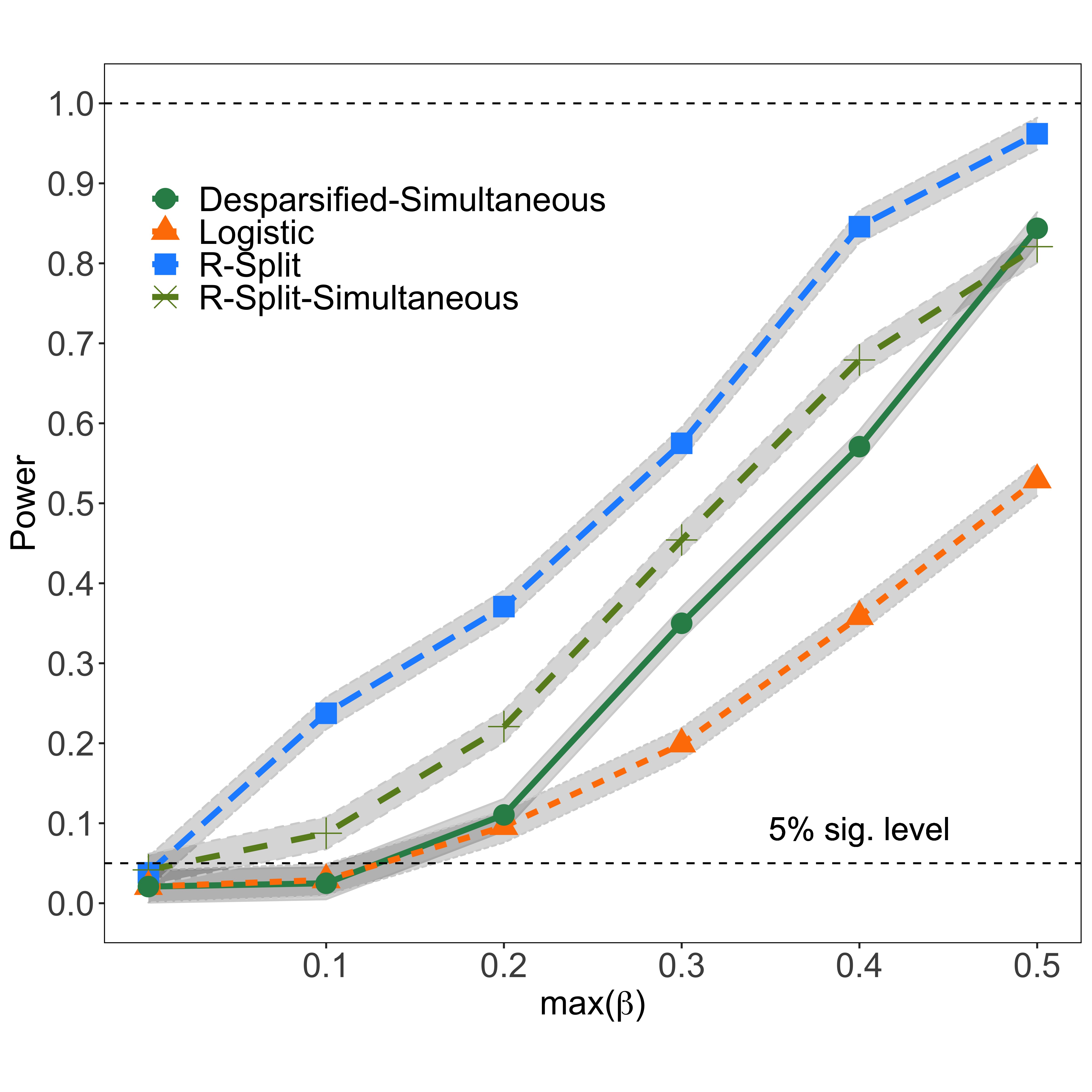}
        \caption{ \label{fig:power}Power comparison for bootstrap-assisted R-Split,  bootstrap-assisted logistic regression, R-Split with simultaneous confidence interval, and the desparsified Lasso with simultaneous confidence interval evaluated over 500 Monte Carlo samples.}
    \end{figure}

The above theoretical result has two direct implications. On the one hand, as the proposed bootstrap calibration strategy successfully replicates the distribution of $\sqrt{n}(\tilde{\beta}_{\max}-\beta_{\max})$, our bias-reduced estimator discussed in the Step 3 of Section \ref{Sec:rsplit-boot} simultaneously removes the regularization bias and the winner's curse bias in $\tilde{\beta}_{\max}$. On the other hand, although simultaneous inference also delivers valid inference on $\beta_{\max}$ with strict Type-I error rate control, our proposal delivers valid inference on $\beta_{\max}$ without sacrificing the statistical power. This property is more desirable in our problem setup as we aim to look for the subgroup with the most severe side effect of statin usage while simultaneous methods often lead to overly conservative conclusions for this purpose. 

To further demonstrate the merit of constructing an asymptotically sharp confidence interval for $\beta_{\max}$ and the benefit of conducting variable selection in finite samples, we compare statistical power for testing the null hypothesis $H_0:\beta_{\max}=0$ for four procedures: (1) the proposed bootstrap-assisted R-Split, (2) R-Split with simultaneous confidence intervals, (3) the proposed bootstrap-assisted logistic regression, and (4) the desparsified Lasso estimator discussed in \cite{zhang2014confidence} with simultaneous confidence interval \citep{dezeure2017high,fuentes2018confidence}. We follow the same simulation setup as the first simulation setup in Example \ref{example:bias}. The tuning parameter is fixed at $r=0.15$ for simplicity. For R-Split, we choose the model size via cross-validation (see Supplementary Materials Section B) with a minimal model size equals 3 and a maximal model size equals 10.

From Figure \ref{fig:power}, we observe that all considered approaches control the Type-I error rate at the nominal level when $\beta_{\max} = 0$. The bootstrap-assisted R-Split has the highest detection power over a range of $\beta_{\max}$ among all considered procedures. The bootstrap-assisted logistic regression has the lowest detection power, which demonstrates the necessity of conducting variable selection to screen out irrelevant predictors. As we have expected, both the R-Split method with simultaneous confidence interval and the desparsified Lasso with simultaneous confidence interval do not retain sufficient statistical power to detect subgroup treatment effect heterogeneity.

\subsection{Simulation studies}\label{Sec:simulation-studies}
	
In this section, we consider various simulation designs to demonstrate the merit of our proposal. There are three main takeaways from this simulation. First, our proposed bootstrap calibration procedure provides confidence intervals with nominal coverage probabilities of $\beta_{\max}$ in finite samples. Second, R-Split based methods provide more accurate point estimates and shorter confidence intervals than the logistic regression based approaches without variable selection. {Third}, the bootstrap-assisted methods have higher statistical efficiency (shorter confidence intervals) compared to the simultaneous methods.

We generate Monte Carlo samples from the following model:
\begin{align*}
 \text{logit}\left\{\mathbb{P}(\y_i=1\mid \z_i,\x_i)\right\} = \z_i\transpose\beta + \x_i\transpose\gamma, \quad i = 1,\ldots,n,
\end{align*}
with $n = 2,000$. We consider two cases for $\beta$: (1) heterogeneous case with $\beta = (0,\ldots, 0, 1)\transpose\in \mathbb{R}^{p_1}$, meaning that there exists subgroup treatment effect heterogeneity and only one subgroup singles out; and (2) spurious heterogeneous case with $\beta = (0,\ldots, 0, 0)\in \mathbb{R}^{p_1}$, meaning that there is no subgroup with significant treatment effect in the population. 
We set $\gamma= (1,1,1,1,0,\dots,0)\in\mathds{R}^{p_2}$.
In all considered simulation designs, we set $p_1\in\{4, 10\}$. We consider the case with $(n, p_2) = (2,000, ~150)$ for logistic regression, R-Split, and the desparsified Lasso \citep{zhang2014confidence},  and consider the case with $(n, p_2) = (2,000, ~500)$ for R-Split and the desparsified Lasso, since logistic regression tends to provide inconsistent estimates in moderately high dimensions \citep{sur2019glm}. In each simulation design, we first take the maximum of estimated subgroup treatment effects, i.e. $\hat{\beta}_{\max} = \max_{j=1,\ldots,p_1}\hat{\beta}_{j}$, in each Monte Carlo sample to mimic the subgroup selection procedure adopted in practice,  and then we take the average across different Monte Carlo samples to calculate the winner's curse bias.

As for the covariates design, we generate $\z_i$ and $\x_i$ from 
\begin{align*}
 \z_{ij} \sim \text{Bernoulli}\Big( \frac{\exp(\x_{i, 2j-1}+ \x_{i, 2j})}{1+\exp(\x_{i, 2j-1}+ \x_{i, 2j})} \Big),  \quad j =1, \ldots, p_1, 
\end{align*} 
where $\x_i\sim N(0,\Sigma) $ with $\Sigma_{ij} = 0.5^{|i-j|}$.
We compare the finite sample performance of the proposed bootstrap-assisted R-Split and the bootstrap-assisted logistic regression with two benchmark methods: (1) a naive method with no bootstrap calibration, which directly uses the estimated maximum coefficient to estimate $\beta_{\max}$ and (2) the simultaneous method as discussed in \cite{dezeure2017high} and \cite{fuentes2018confidence}. For the desparsified Lasso \citep{zhang2014confidence}, we only consider the above-mentioned two benchmark methods: the naive method and the simultaneous method (without bootstrap calibration). For the R-Split method, we choose the model size via cross-validated GLM Lasso with a minimal model size equals 3. We report the coverage probability, the $\sqrt{n}$ scaled confidence interval length and the $\sqrt{n}$ scaled Monte Carlo bias along with their standard errors based on 1,000 Monte Carlo samples in Table \ref{table:hetero}.

Comparing the bootstrap-assisted methods with the naive methods, we observe that the bootstrap-assisted methods have nominal-level coverage, while the naive methods are biased and under-covered. This comparison verifies the theoretical results in Section \ref{subsec:theory-power} that the proposed bootstrap calibration successfully reduces the winner's curse bias.

Comparing the bootstrap-assisted methods with the simultaneous methods, we find that although simultaneous methods have higher coverage probabilities, the confidence intervals are rather long, implying that simultaneous methods are overly conservative.  While our proposed inferential framework reaches the nominal-level coverage probabilities and has shorter confidence intervals leading to asymptotically sharp inference.

The comparison between the bootstrap-assisted R-Split with the bootstrap-assisted logistic regression shows that the latter has larger biases and lower coverage probabilities. The bootstrap-assisted logistic regression has undesirable performance because logistic regression yields biased estimates in moderately high dimensions \citep{sur2019glm}. This comparison reveals the benefit of conducting variable selection when $\gamma$ is sparse and the dimension of covariates is large, and it confirms that R-Split alleviates the regularization bias issue.
Comparing R-Split with the desparsified Lasso, in line with our earlier conjecture in Section \ref{subsec:theory-power}, we observe that the desparsified Lasso approach has wider confidence intervals than those obtained by R-Split and tends to provide conservative inference. 


This simulation study verifies that our proposed inferential framework not only achieves nominal coverage probabilities, but also mitigates the regularization and winner's curse biases. Thus, the proposed inferential framework is sensible to consider for our case study.

\begin{table}[h!]
\caption{ Simulation results (heterogeneous and spurious heterogeneous cases)}\label{table:hetero}

\begin{adjustbox}{width=\textwidth,center}
\small{\begin{tabular}{ccccccccc}
     \hline\hline
           & &\multicolumn{3}{c}{  $\beta=(0,\dots,0,1)\in \mathbb{R}^{p_1}$ (heterogeneity)}& \multicolumn{3}{c}{  $\beta=(0,\dots,0,0)\in \mathbb{R}^{p_1}$ (spurious heterogeneity)} \\
           \cline{2-8}
           &&\multicolumn{3}{c}{Logistic Regression ($\bm{p_2=150}$)} &  \multicolumn{3}{c}{Logistic Regression ($\bm{p_2=150}$)}  \\
             & &  Boot-Calibrated & No adjustment & Simultaneous   &  Boot-Calibrated & No adjustment & Simultaneous\\
       \cline{2-8} 
    $p_1=4$ &Cover  & 0.95(0.01) & 0.89(0.01)& 0.99(0.01)   &0.94(0.02) & 0.87(0.03) &0.99(0.01) & 
    \\  
    
    & $\sqrt{n}$Length & 9.57(0.05) & 8.83(0.03)& 14.6(0.04) &   7.90(0.05)&  6.21(0.05)& 11.1(0.04)  & 
      \\
    
    & $\sqrt{n}$Bias &-2.51(2.70)  &4.91(4.60)&--- &   3.10(3.44) &5.05(4.46) &--- &
    \\[0.15cm]

    $p_1=10$ &Cover & 0.93(0.01) & 0.86(0.01)& 0.99(0.01) & 0.91(0.01) &0.83(0.02) &0.98(0.01) & \\
    
     &$\sqrt{n}$Length & 10.4(0.04)  &9.18(0.04) & 16.7(0.03) & 8.38(0.06) &7.25(0.05) &12.4(0.06) &\\ 
     
     & $\sqrt{n}$Bias &-4.07(3.89)  &5.17(4.67) &--- & 5.30(4.88) &7.32(6.58) &---&
     \\[0.15cm]
      \cline{2-8}

     & & \multicolumn{3}{c}{Repeated Sample Splitting ($\bm{p_2=150}$)} &  \multicolumn{3}{c}{Repeated Sample Splitting ($\bm{p_2=150}$)} \\
     
     &  & Boot-Calibrated & No adjustment & Simultaneous & Boot-Calibrated & No adjustment & Simultaneous & \\  
      \cline{2-8}
     $p_1=4$ &Cover & 0.96(0.01) & 0.94(0.01) &0.99(0.00)&   0.95(0.02) & 0.93(0.02) &0.98(0.02)&  
     \\
      
      &$\sqrt{n}$ Length & 3.56(0.07) &2.17(0.06) &5.14(0.07)  & 1.87(0.04) & 1.03(0.06)&5.08(0.04) &\\
      
       & $\sqrt{n}$Bias & 0.11(0.26) &0.14(0.25) &--- & 0.15(0.24) &0.31(0.39)&--- &
       \\[0.15cm]
       
       $p_1=10$ &Cover & 0.95(0.02) &0.92(0.01) &0.99(0.01)&  0.95(0.01) & 0.91(0.02) &0.96(0.01)&  \\
       
       &$\sqrt{n}$Length &3.62(0.07) &2.57(0.05) &6.61(0.05)&   2.02(0.06)&1.47(0.06) & 6.46(0.04)&  \\
       
        & $\sqrt{n}$Bias & 0.25(0.39) &0.32(0.30) &--- & 0.29(0.40) &0.98(0.90) & --- &\\
        
       \cline{2-8} 
       
       & & \multicolumn{3}{c}{Desparsified Lasso ($\bm{p_2=150}$)} & \multicolumn{3}{c}{Desparsified Lasso ($\bm{p_2=150}$)}  \\
     &  & Boot-Calibrated & No adjustment & Simultaneous & Boot-Calibrated & No adjustment & Simultaneous &  \\  
     \cline{2-8}
   $p_1=4$ &Cover &---  & 0.92(0.01) & 0.99(0.00) &  --- &  0.92(0.01) &0.99(0.01)&  \\ 
&$\sqrt{n}$Length &---   &  2.13(0.06)& 6.51(0.05) & --- & 1.01(0.07)&5.52(0.05)  \\

   & $\sqrt{n}$Bias & ---  & 0.29(0.22)& --- &--- & 1.23(0.99) &--- & 
    \\[0.15cm]
       $p_1=10$ &Cover & ---  & 0.93(0.01)& 0.99(0.01) & --- & 0.93(0.01) &0.99(0.01)& \\
       
       &$\sqrt{n}$Length & ---   & 2.10(0.07)& 6.98(0.07) & --- &1.39(0.06)&6.20(0.07)&    \\
       
         & $\sqrt{n}$Bias &---  & 0.27(0.17) &--- & --- &0.97(0.85) &--- & \\
      \cline{2-8}

& & \multicolumn{3}{c}{Repeated Sample Splitting ($\bm{p_2=500}$)} &  \multicolumn{3}{c}{Repeated Sample Splitting ($\bm{p_2=500}$)} \\
     &  & Boot-Calibrated & No adjustment & Simultaneous & Boot-Calibrated & No adjustment & Simultaneous &  \\  
     \cline{2-8}
   $p_1=4$ &Cover &0.95(0.02) & 0.92(0.03) & 0.99(0.00) & 0.95(0.02) & 0.91(0.02)& 0.98(0.01) &\\ 
&$\sqrt{n}$Length &  4.44(0.06) &  2.22(0.06)& 6.08(0.05) & 3.77(0.05) & 3.18(0.04)& 5.90(0.04) & \\

   & $\sqrt{n}$Bias & -0.68(0.80) & 1.22(1.18)& --- & 0.62(0.72) & 1.58(1.41) & ---   &
    \\[0.15cm]
       $p_1=10$ &Cover & 0.93(0.02) & 0.88(0.03)& 0.98(0.01) & 0.92(0.02) & 0.85(0.01)& 0.95(0.01) & \\
       
       &$\sqrt{n}$Length &  5.11(0.04) & 2.95(0.05)& 6.77(0.05) & 3.54(0.06) & 2.72(0.06)& 6.52(0.05) &   \\
       
         & $\sqrt{n}$Bias &-0.90(0.85) & 1.36(1.20) &--- & 1.53(1.39)& 2.82(1.97)  & ---  &\\
         
      \cline{2-8}

& & \multicolumn{3}{c}{Desparsified Lasso ($\bm{p_2=500}$)} & \multicolumn{3}{c}{Desparsified Lasso ($\bm{p_2=500}$)} &   \\
     &  & Boot-Calibrated & No adjustment & Simultaneous & Boot-Calibrated & No adjustment & Simultaneous &  \\  
     \cline{2-8}
   $p_1=4$ &Cover &---  & 0.90(0.01) & 0.99(0.00) & --- & 0.89(0.01)  &0.99(0.01)& \\ 
&$\sqrt{n}$Length & ---  &  2.19(0.05)& 7.48(0.08) & --- & 3.10(0.05)&6.88(0.08)  & \\

   & $\sqrt{n}$Bias & ---  & 1.29(1.13)& --- & --- &2.30(1.90) &--- & 
    \\[0.15cm]
       $p_1=10$ &Cover & ---  & 0.91(0.01)& 0.99(0.01) & --- & 0.90(0.01) &0.99(0.01)&  \\
       
       &$\sqrt{n}$Length & ---   & 2.15(0.06)& 7.60(0.08) & --- & 2.68(0.05)&7.63(0.08)&  \\
       
         & $\sqrt{n}$Bias & --- & 1.25(1.17) &--- & --- & 2.08(1.96)&--- & \\

     \hline\hline
     \end{tabular}
     }
     \end{adjustbox}
      \begin{tablenotes}\small
   \item Note: ``Cover" is the empirical coverage of the 95\% lower bound for $\beta_{\max}$.  `` $\sqrt{n}$Bias " captures the root-$n$ scaled Monte Carlo bias for estimating $\beta_{\max}$, and  `` $\sqrt{n}$Length " denotes the  root-$n$ scaled length of the 95\% lower bound for $\beta_{\max}$. 
     \end{tablenotes}
 \end{table}

\section{Case study}\label{sec:real-data}

\subsection{Case study results}

In this section, we investigate the adverse effect of statin usage in our pre-specified six subgroups divided by sex and T2D genetic risk using the data introduced in Section \ref{Sec:data-description}. We compare the results from three methods: (1) repeated sample splitting (R-Split) without bootstrap calibration, (2) R-Split based on the simultaneous method discussed in \cite{dezeure2017high},  and (3) the proposed bootstrap-assisted R-Split. We summarize our real data analyses results in Table \ref{table:real-data-two-sided}, in which we have reported the estimated subgroup treatment effects from R-Split along with their $p$-values and two-sided confidence intervals, 
adjusted $p$-values to account for the multiple comparisons issue with simultaneous method and Bonferroni correction, and bootstrap calibrated $p-$values for the subgroup with the largest treatment effect. The results with one-sided confidence lower bounds are summarized in Supplementary Materials Section F.

\begin{table}[h!]
    \centering
\begin{adjustbox}{width={\textwidth},center}%
 \begin{tabular}{ccccc}
    \hline
     \hline\\[-2ex] 
       Method & Subgroup (prevalence; \# of case) & Est (95\% CI)  & $p$-value & Bonf $p$-value  \\
  \\[-2ex] 
    \hline
      \\[-2ex] 
   R-Split & High-risk female $(0.14,100)$  & $0.41~(0.04,~0.78)$ & $0.030$  & $0.180$  \\
    \\[-2ex] 
   (without bootstrap calibration) &Mid-risk female $(0.12,396)$ &  $0.10~(-0.03,~0.24)$ & $0.132$ & $0.792$ \\
  \\[-2ex] 
  &Low-risk female $(0.11,630)$ & $-0.00~(-0.10,~0.09)$ & $0.990$ & $1$ \\
  \\[-2ex]
  & High-risk male $(0.24,139)$  & $-0.07~(-0.38,~0.25)$ & $0.658$  & $1$  \\
    \\[-2ex] 
  &Mid-risk male $(0.21,561)$ & $0.02~(-0.07,~0.11)$ & $0.673$ & $1$ \\
  \\[-2ex] 
  &Low-risk male $(0.17,739)$  &  $-0.03~(-0.16,~0.10)$ & $0.651$ & $1$\\
  \\[-2ex] 
    &Overall  & $0.07~(-0.16,~0.39)$ & $0.545$  & --\\
    \\[-2ex] 
        
         \\[-2ex] 
         Simultaneous & High-risk female $(0.14,100)$ & -- &  $0.256$ & --\\
          \\[-2ex] 
         \hline
         \\[-2ex] 
    Bootstrap-assisted R-Split & High-risk female $(0.14,100)$ & $0.35~(0.02,~0.70)$ &  $0.037$ & --\\
          \\[-2ex] 
     \hline\hline
    \end{tabular}
    \end{adjustbox}
    \caption{\label{table:real-data-two-sided} Estimated treatment effects (Est) on the PHS cohort in six subgroups divided by sex and T2D genetic risk, together with two-sided $95\%$ confidence intervals (CI), corresponding two-sided $p$-values and the Bonferroni $p$-values in the last column. We also present the prevalence of T2D in each subgroup.}
\end{table}

From Table \ref{table:real-data-two-sided}, the results of the R-Split estimator without bootstrap calibration not only indicate that the treatment effect of statins tends to vary across different subgroups, but also suggest that the high-genetic-risk female subgroup is the most vulnerable group for developing T2D with estimated log-odds ratio $0.41$, 95\% two-sided confidence interval $0.04 - 0.78$ (${\rm OR}=1.04 - 2.18$) with $p$-value $0.030$. 
For males with various genetic risk levels and females with lower T2D genetic risk, the adverse effects of statin usage are not significant based on R-Split without bootstrap calibration. The treatment effect in the overall study cohort is slightly positive but is not significant, which is in-line with our expectation from the preliminary analysis in Section \ref{Sec:data-description}.

Although the estimates and confidence intervals from the R-Split without bootstrap calibration suggest that taking statins causes the increased risk of  developing T2D for the most vulnerable subgroup, the statistical significance of this finding is unclear since R-Split is implemented without bootstrap calibration and can not address the multiple comparisons issue as illustrated in Section \ref{Sec:simulation-studies}. After accounting for the multiple comparisons issue through conservative procedures including the simultaneous method or Bonferroni correction, the $p$-values for the female high-risk group are no longer significant, seemingly suggesting that our data do not provide enough evidence to claim the existence of the adverse effect of statin usage in the female high-risk subgroup. This might be due to the fact that both the simultaneous method and Bonferroni correction are rather conservative and tend to provide false negative discoveries. Fortunately, our proposed bootstrap assisted R-Split procedure directly conducts inference on the most vulnerable group, and our results suggest that among high-genetic-risk female patients, the odds of developing T2D after taking statins are 1.42 times the odds of developing T2D for the patients without taking statins ($p-$value $0.037$ for two-sided test).

Our findings are in-line with reported results in existing clinical studies. For example, \cite{mora2010statins} suggest that statin usage incurs a larger T2D risk increment on females than on males, and \cite{waters2013cardiovascular} suggest that statins only significantly increase the risk of T2D on those with at least three out of four common T2D risk factors at baseline.\footnote{The risk factors used by \cite{waters2013cardiovascular} include high fasting blood glucose, history of hypertension, high body mass index, and high fasting triglycerides.} Compared with the existing studies, our findings 
provide more robust evidence with the new data analysis pipeline built under the causal inference framework. Our data analysis pipeline addresses several limitations of existing studies; in particular, limited sample size and multiple comparisons issue. Moreover, 
compared to existing studies, our findings provide a more biologically driven depiction of statins' heterogeneous adverse effect, which can further support effective and precise clinical decisions and actions concerning the prescription of statins. Our study further demonstrates that in practice, the genetic profiles could assist T2D prevention of statin receivers to improve the quality of clinical practices.

\subsection{Sensitivity analysis}

A major concern in observational studies is the bias induced by unmeasured confounding, meaning that some unmeasured factors that are associated with both the treatment and the outcome may explain away the estimated causal effects \citep{robins2000sensitivity}. 
To evaluate the validity of causal conclusions derived from our real data analyses, we conduct sensitivity analyses with the E-value method. The E-value method computes the minimal strength of an unmeasured confounder needed to explain away the estimated causal effect \citep{vanderweele2017sensitivity}. Practitioners could then evaluate if there exists such an unmeasured confounder with the strength quantified by the E-value. A larger E-value implies that the unmeasured confounder needs to have a stronger association with the outcome and the treatment in order to explain away the causal evidence. The E-values for our estimated subgroup causal effects are summarized in Table \ref{table:real-data-sensitivity}. Table \ref{table:real-data-sensitivity} shows that the E-value in the high-risk female group is 2.38, which implies 
that only when an unmeasured confounder is associated with both the treatment and the outcome 2.38 times stronger than the  measured confounders  could the estimated causal effect be explained away. According to a meta-study on E-value applications, most computed E-values from existing literature are below $2.0$ \citep{ioannidis2019limitations}. In sum, the results from Table \ref{table:real-data-sensitivity} imply that the causal evidence collected from our data is reasonably robust against the unmeasured confounding issues. 

\begin{table}[h!]
    \centering
\begin{adjustbox}{width={0.8\textwidth},center}%
 \begin{tabular}{ccccc}
    \hline
     \hline\\[-2ex] 
       Method & Subgroup (prevalence; \# of case) & E-value  &  &    \\
  \\[-2ex] 
    \hline
      \\[-2ex] 
   R-Split & High-risk female $(0.14,100)$  & $2.38$ &  &  \\
    \\[-2ex] 
   (without bootstrap calibration) &Mid-risk female $(0.12,396)$ &  $1.45$ &  &  \\
  \\[-2ex] 
  &Low-risk female $(0.11,630)$ & $1.00$ & &  \\
  \\[-2ex]
  & High-risk male $(0.24,139)$  & $1.23$ &  &   \\
    \\[-2ex] 
  &Mid-risk male $(0.21,561)$ & $1.11$ &  &  \\
  \\[-2ex] 
  &Low-risk male $(0.17,739)$  &  $1.14$ &  &\\
  \\[-2ex] 
    &Overall  & $1.23$ &   & \\
    \\[-2ex] 
        
          \\[-2ex] 
         \hline
         \\[-2ex] 
    Bootstrap-assisted R-Split & High-risk female $(0.14,100)$ & $2.19$ &  & \\
          \\[-2ex] 
     \hline\hline
    \end{tabular}
    \end{adjustbox}
    \caption{\label{table:real-data-sensitivity} Sensitivity analysis of our causal evidence measured by the E-value. }
\end{table}

Given that our outcome is an error-prone surrogate of the true disease status,  we also conduct a sensitivity analysis regarding the potential misspecification of the logistic regression model for the true EHR disease status against the covariates. Due to page limit, the design and results of this sensitivity analysis are deferred to Supplementary Materials Section I.

\section{Discussion}
In this case study, we investigate the T2D risk associated with statin usage in the most vulnerable subgroup. To overcome the limitations of existing studies and to generate trustworthy evidence, we introduce a rigorous study design under the causal inference framework and based on the EHR and biobank data from the Partner Health System. Built on this study design, we find that although the adverse effect of statin usage for developing T2D is marginal for the overall study cohort, taking statins significantly increases the risk of developing T2D for female patients with high genetic predisposition to T2D. We also recognize that our study design has two limitations. First, as the treatment variable is defined as if the subject carries the rs12916-T allele or not, we can only investigate the causal effect of taking statins on T2D risk but not the \textit{dosage effect} of statins. Second, the definition of T2D status is based on a previously validated Multimodal Automated Phenotyping (MAP) algorithm \citep{liao2019high}. Although  the MAP classifier of T2D can be reliably used to define the T2D outcome, generalizing the current study findings still warrants further confirmation from clinical trials.

While the objective of this case study is to make inference on the most vulnerable subgroup, a natural question to ask is whether statin usage will significantly increase the T2D risk for other vulnerable subgroups. To answer this question, we need to develop appropriate statistical tools to mitigate the regularization bias and winner's curse bias for other most vulnerable subgroups as well. Take the subgroup with the second largest treatment effect as an example, our proposed method might be extended to address the bias issues by appropriately modifying the correction term $\tilde{c}_j(r)$ to capture the distance between the second largest coefficient and the $j$-th largest coefficient.  
We shall leave the rigorous methodology development for making valid inference on the other subgroups to future research.


This case study considers pre-defined candidate subgroups. While predefined subgroups are suitable in our case study (as discussed in Section \ref{subsec:study-design}), extending the proposed methodology to data-adaptively identified subgroups warrants future research. Data-adaptive subgroup identification approaches include, for example, varying coefficient model based \citep{chen2018inference}, regression tree based \citep{lipkovich2011subgroup}, and fused Lasso based \citep{ma2017concave} methods. When working with data-adaptively identified subgroups, one needs to not only adjust for the regularization and winner's curse bias, but also account for randomness induced by subgroup identification. 
We leave this possible extension of the proposed method for future research, as the primary objective of this manuscript is to investigate the causal effect of statin usage on T2D risk in the most vulnerable subgroup.

\subsection*{Software and reproducibility}
\texttt{R} code for the proposed procedures can be found in the package ``\texttt{debiased.subgroup}'' that is publicly available at \texttt{https://github.com/WaverlyWei/debiased.subgroup}. Simulation examples can be reproduced by running examples in the \texttt{R} package.

\subsection*{Acknowledgement}
The authors would like to thank the editor, the associate editor, and anonymous reviewers for their comments and suggestions that significantly improved the paper. The authors also thank Xuming He and Xinwei Ma for their valuable feedback and thoughtful discussions.

\clearpage

\bibliographystyle{jasa} 
	\bibliography{reference}

	\clearpage

\appendix
\addcontentsline{toc}{section}{Appendix} 
\part{Appendix} 
\parttoc 
\onehalfspacing
\setcounter{page}{1}

\section{Refitting bias and selection bias}\label{appendix:bias-issues}

In this section, we demonstrate the refitting bias issues discussed in the main manuscript with illustrative derivations. To facilitate discussion, suppose for now that $f_i \triangleq \text{expit}'(\z_i^\intercal\beta + \x_i^\intercal\gamma)$ is given. Note that this is infeasible in practice, and neither our theoretical investigation nor our practical implementation requires $f_i$ to be known. 

We start with some illustrative derivations on the regularization bias. Since $\beta\in\mathbb{R}^{p_1}$ is a low-dimensional parameter of interest, penalizing $\beta$ is not necessary in our problem setup. Instead, inference on $\beta$ can be carried out after a small number of predictors in $\mbox{x}$ are selected \citep{belloni2013least, belloni2014inference}. We use $\hm$ to record this selected set of predictors. Then, the refitted GLM estimator is obtained via minimizing the negative log-likelihood function
	 \begin{align*}
	 (\hat{\beta}_{\text{GLM}}\transpose, \hat{\gamma}_{\text{GLM}}\transpose)\transpose =\underset{\beta\in \mathbb{R}^{p_1}, \gamma\in \mathbb{R}^{|\hm|}} {\arg\min} \left\{ \frac{1}{n}\sum_{i=1}^n \Big(\y_i\cdot( \z_i^\intercal\beta + \x_{i, \hm}^\intercal\gamma)-\log \big(1+\exp(\z_i^\intercal\beta +\x_{i, \hm}^\intercal\gamma)\big)\Big)\right\}.
	 \end{align*}
Under the impact of the random model $\hm$ entering the estimation process, $\hat{\beta}_{\text{GLM}}$ often cannot consistently estimate $\beta$ unless perfect model selection is achieved (i.e., $\hm = \m_0$). To see this, following the derivation provided in the appendix (Section B), we can decompose $\hat{\beta}_{\text{GLM}}$ into two parts: 
	\begin{align}\label{eq:refitting-bias-decomposition}
	\sqrt{n}(\hat{\beta}_{\text{GLM}} -\beta)= \underbrace{\mbox{I}_{\z}(\hat{\Sigma}_{\hm})^{-1}\cdot \frac{1}{\sqrt{n}}\sum_{i=1}^n\begin{pmatrix}
	\z_i\\
	\x_{i,\hm}
	\end{pmatrix}\nu_i}_{=:b_{n1} } + \underbrace{\vphantom{\sum_{i=1}^n}(\tilde{\bm{\z}}^\intercal(\bm{\mbox{I}}-\tilde{\bm{\mbox{P}}}_{\hm})\tilde{\bm{\z}}/n)^{-1}\tilde{\bm{\z}}^\intercal(\bm{\mbox{I}}-\tilde{\bm{\mbox{P}}}_{\hm})\tilde{\bm{\x}}\beta/\sqrt{n}}_{=:b_{n2}} + o_p(1),
	\end{align}
where $\nu_i = \y_i - \text{expit}(\z_i\transpose\beta + \x_i^\intercal\gamma)$, 
$\mbox{I}_{\z}$ denotes an index matrix such that $\mbox{I}_{\z}(\z_i\transpose,\x\transpose_{i,\hm})\transpose = \z_i\transpose$, and $\hat{\Sigma}_{\hm} $ is the sample Hessian matrix defined as
$\hat{\Sigma}_{\hm} = \frac{1}{n}\sum_{i=1}^n f_i (\z_i^\intercal,\x^\intercal_{i,\hm})^\intercal(\z_i^\intercal,\x^\intercal_{i,\hm}).$
 Furthermore, $\bm{\mbox{D}}=\text{diag}(\bm{\mbox{D}}) = (f_1, \ldots, f_n)$ is a diagonal matrix.  Then, $\tilde{\bm{\z}}\transpose = \bm{\z}\transpose\bm{\D}^{1/2}$,  $\tilde{\bm{\mbox{x}}}\transpose = \bm{\x}\transpose\bm{\D}^{1/2}$, $\tilde{\bm{\x}}_{\hm}^\intercal = \bm{\x}_{\hm}^\intercal\bm{\D}^{1/2}$, and the projection matrix is $\bm{\tilde{\bm{\mbox{P}}}_{\hm}} = \tilde{\bm{\x}}_{\hm} (\tilde{\bm{\x}}_{\hm}\transpose\tilde{\bm{\x}}_{\hm})^{-1}\tilde{\bm{\x}}_{\hm}\transpose$. 
 
The regularization bias has two sources implied by the decomposition above. The first bias term $b_{n1}$ is often not centered around zero due to the correlation between $\nu_i$ and the data dependent model (i.e., $\mathbb{E}(\nu_i|\x_{i,\hm}) \neq 0$). As this bias only occurs whenever an irrelevant variable is selected. The second term $b_{n2}$ captures the impact of omitting variables in the true model $\texttt{M}_0$ for estimating $\beta$, and it occurs whenever the selected model $\hm$ under-covers the true support set of $\gamma$ (i.e., $\texttt{M}_0$). The impact of the under-fitting vanishes whenever the sure screening property, $\texttt{M}_0 \subseteq \hm $, holds. Existing literature in linear models has argued that sufficient conditions for sure screening property to hold are much weaker than the ones needed for the perfect model selection \citep{wasserman2009high}, indicating selecting a larger model can be a simple remedy to avoid the under-fitting bias.

\section{Implementation details}\label{appendix:implementation}
	\begin{description}
	\item[Step 1.] For $b\leftarrow 1$ to $B_1$ do
	\begin{enumerate}
	    \item Randomly split the data $\{ (\y_i, \x_i, \z_i) \}_{i=1}^n$ into group  $T_1$ of size $n_1$ and group $T_2$ of size $n_2 = n-n_1$, for $i=1,\cdots,n$. 
	    \item  Select a model $ \hm_b$ to predict $\y$  based on $T_1$.
	    \item Refit the model with the data in $T_2$ to get
	    \begin{align*}
	        (\tilde{\beta}_b\transpose  ,\tilde{\gamma}_b\transpose)\transpose  =  \arg\min \left\{ \sum_{l\in T_2}\Big(\y_l\cdot( \z_l\transpose \beta + \x_{l, \hm_b}\transpose \gamma)-\log \big(1+\exp(\z_l\transpose \beta +\x_{l, \hm_b}\transpose \gamma)\big)\Big)\right\}.
	    \end{align*}
	      \item Let $f_{bl} = \text{expit}'(\z_l\transpose \tilde{\beta}_b + \x\transpose _{l,\hm_b}\tilde{\gamma}_b)$.
	    \item The R-Split estimate is obtained by averaging over $\tilde{\beta}_b$: 
	\begin{align*}
	    \widetilde{\beta} = \frac{1}{B_1} \sum_{b=1}^{B_1} \tilde{\beta}_b.
	\end{align*}
	\end{enumerate}
	
	\item[Step 2.] 
	\begin{enumerate}
	   
	    \item For $j\in [p_1]$, calculate: 
	$$\tilde{\Gamma}_n = \frac{1}{B_1}\sum_{b=1}^{B_1}    \mbox{I}_z \left( \frac{1}{n_1}\sum_{i=1}^n \mathbf{1}_{(i\in T_2)}\cdot f_{bl} \begin{pmatrix} \z_i \\
			\x_{i,\hm} \end{pmatrix} (\z_i\transpose , \x\transpose _{i,\hm})\transpose  \right)^{-1}\mbox{I}_{\hm},\quad \tilde{c}_j(r) = (1-n^{r-0.5})(\tilde{\beta}_{\max} - \tilde{\beta}_j),$$
		where $r$ is a positive tuning parameter between 0 to 0.5. 
	\end{enumerate}
	\item[Step 3.] For $b\leftarrow 1$ to $B_2$ do
	\begin{enumerate}
	    \item Generate bootstrap replicate $\tilde{\beta}^*$:
	    \begin{align*}
    \tilde{\beta}^* =  \tilde{\beta}+\tilde{\Gamma}_n \cdot \frac{1}{n} \sum_{i=1}^n \begin{pmatrix}
    \z_i\\
    \x_{i}
    \end{pmatrix} \nu^*_i.
    \end{align*}
	    \item Recalibrate bootstrap statistics  via
	    \begin{align*}
	        T_b^* =\underset{j\in [p_1]}{\max}(\tilde{\beta}^*_{j}+\tilde{c}_j(r))- \tilde{\beta}_{\max}.
	    \end{align*}
	\end{enumerate}
	\item[Step 4.]  The level-$\alpha$ one-sided confidence interval for $\beta_{\max}$ is  $[\tilde{\beta}_{\max}-{Q}_{T^*_b}(\alpha),+\infty)$, and the level-$\alpha$ two-sided confidence interval for $\beta_{\max}$ is  $[\tilde{\beta}_{\max}-{Q}_{T^*_b}(\alpha),\tilde{\beta}_{\max}+{Q}_{T^*_b}(\alpha)]$ and a bias-reduced estimate. 
	\end{description}


In Step 1 (1), we recommend a split ratio of $0.6:0.4$ for $n_1:n_2$ because a larger sample size for subsample $T_1$ improves model selection accuracy. In Step 1 (2), the model selection procedure can be any easily accessible procedure. In our case, since the real data have binary outcomes, we adopt GLM lasso for model selection with R package \texttt{glmnet} \citep{hastie2007glmlasso}. The model size is selected via cross-validation with a constraint on the maximal and minimal model sizes. We recommend to set $B_1 = 500$ for the number of repeated splits and $B_2 = 1,000$ for the number of bootstrap replications. As for the tuning parameter $r$, we propose a data-adaptive cross-validated algorithm to select $r$ as the following  \citep{guo2020inference}:

\begin{description}\label{algorithm:tuning}
    \item[Step 1] \textit{Denote $R=\{r_1,\dots,r_m\}$  as a set of candidate tuning parameters. Randomly split the sample into $v$ equal-sized subsamples.} 
    \item[Step 2]\textit{For l $\leftarrow$ 1 to m:}
\item[\quad\quad\quad\quad] \textit{For j $\leftarrow$ 1 to $v$:}
\item[\quad\quad\quad\quad\quad\quad (a)] \textit{Use subsample $j$ as reference data and the rest as training data. Obtain \makebox[2.5cm]{\hfill}  $\tilde{\beta}_{\max,\mathrm{reduced},j}(r_l)$ on the training data, where $r_l$ is the tuning parameter.}
\item[\quad\quad\quad\quad\quad\quad (b)] \textit{For i $\leftarrow$ 1 to k: Obtain R-Split estimate of $\tilde{\beta}_{i,j}$ and its standard error \makebox[2.5cm]{\hfill} $\tilde{\sigma}_{i,j}$ on the reference data; evaluate the accuracy
\begin{align*}
    h_{i,j}(r_l)=(\tilde{\beta}_{\max,\mathrm{reduced},j}(r_l)-\tilde{\beta}_{i,j})^2-\tilde{\sigma}_{i,j}^2.
\end{align*}}
\item[Step 3]\textit{Select the tuning parameter via $\arg\min_{r_l} \{ \min_{i\in [k]}[\sum_{j=1}^{j=v}h_{i,j}(r_l)/v] \}.$}
\end{description}

Intuitively, we want to choose $r$ that minimizes the mean squared error between the proposed bias reduced estimate $\tilde{\beta}_{\max,\mathrm{reduced}}$ and $\beta_{\max}$. Because $\beta_{\max}$ is unknown, we provide an approximation of the mean squared error that can be computed via cross-validation in Step 2 (b). The justification of this cross validation method for fixed $p_1$ can be found in \cite{guo2020inference}. In our empirical work, we implement the above tuning selection method via three-fold cross-validation with a candidate set $R = \{1/3,1/6,\dots,1/30\}$.

\section{Theoretical investigation for R-Split assisted bootstrap calibration}\label{appendix:rsplit-theory}

In this section, first, we show the asymptotic consistency of R-Split estimator under generalized linear models. Second, we show the bootstrap consistency result of R-Split with fixed $p_1$. Lastly, we provide theoretical details of Theorem 1. Our proofs rely on the following assumptions.

\subsection{Assumptions}\label{appendix:assumption}
\begin{assumption}{1}{} \label{assumption:data}
 Suppose $\{(\y_i,\z_i,\x_i)\transpose\}_{i=1}^n$ is a random sample and $(\z_i,\x_i)$ have zero mean and bounded support with an upper bound $C$, i.e. $|\z_{ij}|\leq C$, $|\x_{ij}|\leq C$, $|\x_{ij}\x_{ij}\transpose|\leq C$, for $i=1,\ldots,n, j = 1,\ldots,p$. 
 \end{assumption}

 \begin{assumption}{2}{} \label{assumption:split}
 The split ratio $\mbox{r} = n_2/n$ is a constant in $(0,1)$. The selected model sizes in all splits are bounded by $\mbox{S}$ with $\mbox{S} = o(n)$.
 \end{assumption}

  \begin{assumption}{3}{}
  \label{assumption:overfitting}
\begin{align*}
    \frac{1}{\sqrt{n}}\sum_{i=1}^n\Bigg\{\Big\{\mathbb{E}\Big(s_ia_1\transpose\hat{\Sigma}^{-1}_{\hm,\s}\mbox{I}_{\hm}|\y,\z,\x\Big) - \mathbb{E}\Big(\tilde{s}_ia_1\transpose\hat{\Sigma}^{-1}_{\tilde{\m},\tilde{s}}\mbox{I}_{\tilde{\m},\tilde{\s}}|\y,\z,\x\Big)\Big\}(\z\transpose_i,\x\transpose_i)\transpose\nu_i\Bigg\} = o_p(1),
\end{align*}
 \end{assumption}

\begin{assumption}{4}{}
\label{assumption:random-vector}
  For any vector $a_1\in \mathbb{R}^{p_1}$, there exists a random vector $\eta_n(a_1)\in\mathbb{R}^{p+1}$ which is independent of $\nu$, and $||\eta(a_1)_n||_{\infty}$ is bounded in probability and satisfies
  \begin{align*}
      \Big|\Big|\mbox{r}_{\s}\mathbb{E}\Big(a_1\transpose\hat{\Sigma}^{-1}_{\hm,\s}\mbox{I}_{\hm}|\y,\z,\x\Big) - \eta\transpose_n(a_1)\Big|\Big|_1 = o_p(1/\sqrt{\log p}).
  \end{align*}
\end{assumption}

\begin{assumption}{5}{}
\label{assumption: under-fitting}
The under-fitting bias over all splits is negligible, such that
\begin{align*}
     \mathbb{E} \Big(\big(\tilde{\bm{\z}}_{\mbox{s}}\transpose(\bm{\mbox{I}}-\tilde{\bm{\mbox{P}}}_{\hm,\mbox{s}})\tilde{\bm{\z}}_{\s}/n\big)^{-1}\tilde{\bm{\z}}_{\s}^\intercal(\bm{\mbox{I}}-\tilde{\bm{\mbox{P}}}_{\hm,\s})\tilde{\bm{\x}}_{\s}\beta/\sqrt{n}|\y,\z,\x\Big) = o_p(1).
\end{align*}
\end{assumption}

 \begin{assumption}{6}{} \label{assumption:differentiability}
 We assume $\text{expit}(\y|\z,\x)$ is continuously differentiable in $\y$ for each $(\z\transpose,\x\transpose)$ in the support of $(\z\transpose,\x\transpose)$ and $|\text{expit}'(\y|\z,\x)| \leq C$, uniformly in $\y$ and $(\z\transpose,\x\transpose)$. 
 \end{assumption}
 
 \begin{assumption}{7}{}
 \label{assumption: glm-lasso-rate}
\begin{align*}
    \mathbb{E}\Big[\Big(\text{expit}(\z\transpose\hat{\beta} +\x\transpose\hat{\gamma}) - \text{expit}(\z\transpose\beta +\x\transpose\gamma) \Big)^2\Big] = O_p\Big(\frac{|\m_0|\log (p)^{3/2+\delta}}{n}\Big), \quad \delta >0.
\end{align*}
 \end{assumption}

 \begin{assumption}{8}{}
 \label{assumption: eigen}

There exist a constant $U$ and $L$ where $L\leq \tilde{\Sigma}_{n;i,i}\leq U$ for any $i\in [p_1]$ and $\tilde{\Sigma}_{n}^{-1}$ exists where $\tilde{\Sigma}_{n}$ is defined in C.3.

 \end{assumption}
 
  \begin{assumption}{9}{}
 \label{assumption: distinguishable}

$\max_{i\in H}\beta_i-\max_{i\notin H}\beta_i\ge \tilde{L}$ where $\tilde{L}$ is a constant and $H=\{j:\beta_j=\beta_{\max}\}$.

 \end{assumption}

\begin{lem}\label{lemma1}
     Under Assumption 1, $||\bm{\z}\transpose \nu/\sqrt{n}||_{\infty} = O_p(\sqrt{\log p})$
\end{lem}
\begin{proof}
For $K>0$,
\begin{align*}
     &\mathbb{P}\Big(\max_j\Big|\sum_{i=1}^n\z_{ij}\nu_i/\sqrt{n}\Big| > \sqrt{\log p}K\Big),\\
     &\leq \mathbb{E}\Big\{\mathbb{P}\Big(\max_j||\z_j||_2\cdot \max_j \Big|\sum_{i=1}^n \frac{\nu_i\z_{ij}}{||\z_j||_2}\Big| > \sqrt{\log p}K \Big| \bm{\z}\Big)\Big\},\\
     &\leq p\mathbb{E}\Big\{\mathbb{P}\Big(\Big|\sum_{i=1}^n \frac{\nu_i\z_{ij}}{||\z_j||_2}\Big| > \sqrt{\log p}K /\max_j ||\z_j||_2\Big| \bm{\z}\Big)\Big\},\\
     &\leq 2\exp\Big(\log p - \frac{\log p K^2}{2\sigma_{\nu}^2C^2}\Big),
\end{align*}
where the last line is by Hoeffding's inequality. 
\end{proof}
Assumption \ref{assumption:data} applies upper bounds on the covariates. Assumption \ref{assumption:split} puts a constraint on the selected model size. Assumption \ref{assumption:overfitting} implies that conditioning on subsample $S$ or $\tilde{S}$ yields the same distributions.  Assumption \ref{assumption:random-vector} implies $\hat{\Sigma}^{-1}_{\hm}$ converges to a random vector $\eta_n$ with error rate $1/\sqrt{\log p}$. Assumption \ref{assumption: under-fitting} assumes the under-fitting bias is negligible \citep{wang2018debiased}. Assumption \ref{assumption:differentiability} assumes smoothness condition for the $\textit{expit}$ function. Assumption \ref{assumption: glm-lasso-rate} provides the convergence rate of GLM Lasso \citep{farrell2015robust}. Assumption \ref{assumption: eigen} basically requires the variance of $\tilde{\beta}$ is bounded above and below, and Assumption \ref{assumption: distinguishable} requires that the best subgroup is separable from the second best one.

\subsection{Proofs of R-Split's asymptotic normality and bootstrap consistency under GLM}\label{appendix:consistency}	
In this section, first, we prove R-Split's asymptotic consistency and normality under GLM as stated in Theorem \ref{thm:normality-rsplit}. Then we show R-Split's bootstrap consistency in the later part of the section. 

\subsubsection{R-Split's asymptotic and consistency and normality under GLM }
\begin{thm}[Asymptotic normality of R-Split under generalized linear models]\label{thm:normality-rsplit}
Under Assumptions \ref{assumption:data} - \ref{assumption: glm-lasso-rate}, the smoothed estimator from R-Split under GLM is asymptotically consistent, such that
\begin{align*}
    \sqrt{n}a_1\transpose(\tilde{\beta}-\beta) = \eta_n\transpose(a_1)\frac{1}{\sqrt{n}}\sum_{i=1}^n  (\z_i\transpose,\x_i\transpose)\transpose \nu_i + o_p(1),
\end{align*}
where $a_1$ is a random vector, $||a_1||_2 = 1$. $\eta_n(a_1)$ is a random vector as a function of $a_1$, $\nu_i = \y_i - \text{expit}(\z_i\transpose\beta + \x_i\transpose\gamma)$. Let $\tilde{\sigma} = \sigma_{\nu}\big(\eta\transpose_n(a_1)\hat{\Sigma}_n\eta_n(a_1)\big)^{1/2}$, 
\begin{align*}
    \tilde{\sigma}^{-1}\sqrt{n}a_1\transpose(\tilde{\beta}-\beta)\leadsto N(0,1),
\end{align*}
where $\hat{\Sigma}_n = \frac{1}{n}\sum_{i=1}^n f_i(\z_i\transpose, \x_i\transpose)(\z_i\transpose, \x_i\transpose)\transpose$, $f_i =\text{expit}'(\z_i\transpose\beta + \x_i\transpose\gamma)$.
\end{thm}

The proof of Theorem \ref{thm:normality-rsplit} follows three steps: (1) decompose refitting bias, (2) show asymptotic consistency of R-Split under GLM and (3) prove asymptotic normality of R-Split under GLM.

\begin{proof}

\subsubsection*{Step 1. Refitting bias decomposition}
Here, our goal is to provide refitting bias decomposition under GLM to cast some insights on the refitting bias issue and also simplify the later asymptotic consistency proof. We want to show the refitting bias can be decomposed as 
\begin{equation}\label{eqn:decomp}
	\sqrt{n}(\hat{\beta}_{\texttt{GLM}} -\beta)= \mbox{I}_{\z}(\hat{\Sigma}_{\hm})^{-1}\cdot \frac{1}{\sqrt{n}}\sum_{i=1}^n (\z\transpose,\x_{\hm}\transpose)\transpose\nu_i + (\tilde{\bm{\z}}^\intercal(\bm{\mbox{I}}-\tilde{\bm{\mbox{P}}}_{\hm})\tilde{\bm{\z}}/n)^{-1}\tilde{\bm{\z}}^\intercal(\bm{\mbox{I}}-\tilde{\bm{\mbox{P}}}_{\hm})\tilde{\bm{\x}}\beta/\sqrt{n}.
\end{equation}

To start, since the refitted estimator in GLM satisfies: 
	 \begin{align*}
(\hat{\beta}_{\texttt{GLM}}\transpose, \hat{\gamma}\transpose_{\texttt{GLM}})\transpose =\underset{ \substack{ \beta\in \mathbb{R}^{p_1}, \gamma\in \mathbb{R}^{p_2}\\ \gamma_j = 0,\ j\not\in \hm }} {\arg\min} \left\{ \frac{1}{n}\sum_{i=1}^n \Big(\y_i\cdot( \z_i\transpose\beta + \x_{i, \hm}\transpose\gamma)-\log \big(1+\exp(\z_i\transpose\beta +\x_{i, \hm}\transpose\gamma)\big)\Big) \right\},
\end{align*}
they are the solution to the following equation: 
\begin{align*}
\sum_{i=1}^n  \begin{bmatrix}
\z_i\\
\x_{i,\hm}
\end{bmatrix} \cdot \big[ \nu_i + \text{expit}( \z_i\transpose\beta + \x_i\transpose\gamma ) -   \text{expit}( \z_i\transpose\hat{\beta}_{\texttt{GLM}} + \x_{i}\transpose\hat{\gamma}_{\texttt{GLM}})  \big] = 0, 
\end{align*}
where $\nu_i = \y_i -  \text{expit}( \z_i\transpose\beta + \x_i\transpose\gamma ) $ is a mean-zero random variable. By Taylor expansion, without loss of generality we assume that there exists some intermediate vectors $\tilde{\beta}_{\texttt{GLM}} \in ( \beta,  \hat{\beta}_{\texttt{GLM}} )$ and $\tilde{\gamma}_{\texttt{GLM}} \in ( \gamma,  \hat{\gamma}_{\texttt{GLM}} )$ such that 
\begin{align*}
 \text{expit}( \z_i\transpose\beta + \x_i\transpose\gamma ) -   \text{expit}( \z_i\transpose\hat{\beta}_{\texttt{GLM}} + \x_{i}\transpose\hat{\gamma}_{\texttt{GLM}}) = \text{expit}'( \z_i\transpose\tilde{\beta}_{\texttt{GLM}} + \x_{i}\transpose\tilde{\gamma}_{\texttt{GLM}}) \cdot \big[ \z_i\transpose (\beta - \tilde{\beta}_{\texttt{GLM}})  + \x_i\transpose (\gamma - \tilde{\gamma}_{\texttt{GLM}}) \big].
\end{align*}
Thus, by denoting $ \m_1 = \{1, \ldots,p_2\} \backslash \hm $, we have 
\begin{align*}
\sum_{i=1}^n  \begin{bmatrix}
\z_i\\
\x_{i,\hm}
\end{bmatrix} \cdot \nu_i  = &  \sum_{i=1}^n \text{expit}'( \z_i\transpose\tilde{\beta}_{\texttt{GLM}} + \x_{i}\transpose\tilde{\gamma}_{\texttt{GLM}})    \begin{bmatrix}
\z_i\\
\x_{i,\hm}
\end{bmatrix} \cdot \big[ \z_i\transpose (\hat{\beta}_{\texttt{GLM}} - \beta) + \x_i\transpose (\hat{\gamma}_{\texttt{GLM}} - \gamma  ) \big], \\ 
= &  \sum_{i=1}^n \text{expit}'( \z_i\transpose\tilde{\beta}_{\texttt{GLM}} + \x_{i}\transpose\tilde{\gamma}_{\texttt{GLM}})    \begin{bmatrix}
\z_i\\
\x_{i,\hm}
\end{bmatrix} \cdot \big[ \z_i\transpose(  \hat{\beta}_{\texttt{GLM}} -\beta)+ \x_{i, \hm}\transpose (\hat{\gamma}_{\texttt{GLM}, \hm} - \gamma_{\hm}  )  + \x_{i, \m_1 }\transpose  (\hat{\gamma}_{\texttt{GLM}, \m_1} - \gamma_{\m_1}  )\big],  \\ 
= &  \sum_{i=1}^n \text{expit}'( \z_i\transpose \tilde{\beta}_{\texttt{GLM}} + \x_{i}\transpose\tilde{\gamma}_{\texttt{GLM}})    \begin{bmatrix}
\z_i\\
\x_{i,\hm}
\end{bmatrix} \cdot \big[ \z_i\transpose , \x_{i, \hm}\transpose, \x_{i, \m_1}\transpose \big] \begin{bmatrix}
  \hat{\beta}_{\texttt{GLM}} - \beta \\
  \hat{\gamma}_{\texttt{GLM}, \hm} - \gamma_{\hm}  \\
  \hat{\gamma}_{\texttt{GLM}, \m_1} - \gamma_{\m_1} 
\end{bmatrix}, \\
= &  \sum_{i=1}^n \text{expit}'( \z_i\transpose \tilde{\beta}_{\texttt{GLM}} + \x_{i}\transpose\tilde{\gamma}_{\texttt{GLM}})    \begin{bmatrix}
\z_i\\
\x_{i,\hm}
\end{bmatrix} \cdot \big[ \z_i\transpose , \x_{i, \hm}\transpose\big] \begin{bmatrix}
\hat{\beta}_{\texttt{GLM}} - \beta \\
\hat{\gamma}_{\texttt{GLM}, \hm} - \gamma_{\hm}  
\end{bmatrix} \\
&  + \sum_{i=1}^n \text{expit}'( \z_i\transpose \tilde{\beta}_{\texttt{GLM}} + \x_{i}\transpose\tilde{\gamma}_{\texttt{GLM}})    \begin{bmatrix}
\z_i\\
\x_{i,\hm}
\end{bmatrix} \cdot \x_{i, \m_1}\transpose\big[ \hat{\gamma}_{\texttt{GLM}, \m_1} - \gamma_{\m_1} \big].
\end{align*}
By Assumption \ref{assumption: under-fitting}, when $\m_0 \subset \hm$, the second term is $o_p(1)$. Rearranging the first term, we have
\begin{align*}
\begin{bmatrix}
\hat{\beta}_{\texttt{GLM}} - \beta \\
\hat{\gamma}_{\texttt{GLM}, \hm} - \gamma_{\hm}  
\end{bmatrix} = & \left(  \sum_{i=1}^n \text{expit}'( \z_i\transpose \tilde{\beta}_{\texttt{GLM}} + \x_{i}\transpose\tilde{\gamma}_{\texttt{GLM}})    \begin{bmatrix}
\z_i\\
\x_{i,\hm}
\end{bmatrix} \cdot \big[ \z_i\transpose , \x_{i, \hm}\transpose \big] \right)^{-1} \sum_{i=1}^n  \begin{bmatrix}
\z_i\\
\x_{i,\hm}
\end{bmatrix} \cdot \nu_i  \\
& - \left(  \sum_{i=1}^n \text{expit}'( \z_i\transpose \tilde{\beta}_{\texttt{GLM}} + \x_{i}\transpose\tilde{\gamma}_{\texttt{GLM}})    \begin{bmatrix}
\z_i\\
\x_{i,\hm}
\end{bmatrix} \cdot \big[ \z_i\transpose , \x_{i, \hm}\transpose \big] \right)^{-1} \\
& \qquad \cdot \sum_{i=1}^n \text{expit}'( \z_i\transpose \tilde{\beta}_{\texttt{GLM}} + \x_{i}\transpose\tilde{\gamma}_{\texttt{GLM}})    \begin{bmatrix}
\z_i\\
\x_{i,\hm}
\end{bmatrix} \cdot \x_{i, \m_1}\transpose\big[ \hat{\gamma}_{\texttt{GLM}, \m_1} - \gamma_{\m_1} \big].
\end{align*}

Therefore, the refitting bias of GLM estimator $\hat{\beta}$ can be decomposed as:
\begin{equation*}
	\sqrt{n}(\hat{\beta}_{\texttt{GLM}} -\beta)= \mbox{I}_{\z}(\hat{\Sigma}_{\hm})^{-1}\cdot \frac{1}{\sqrt{n}}\sum_{i=1}^n (\z\transpose,\x_{\hm}\transpose)\transpose\nu_i + (\tilde{\bm{\z}}^\intercal(\bm{\mbox{I}}-\tilde{\bm{\mbox{P}}}_{\hm})\tilde{\bm{\z}}/n)^{-1}\tilde{\bm{\z}}^\intercal(\bm{\mbox{I}}-\tilde{\bm{\mbox{P}}}_{\hm})\tilde{\bm{\x}}\beta/\sqrt{n},
\end{equation*}
where $\mbox{I}_{\z}$ denotes an index matrix, $\mbox{I}_{\z}(\z_i\transpose,\x\transpose_{i,\hm})\transpose= \z_i\transpose$. Define the sample Hessian matrix as 
$$\hat{\Sigma}_{\hm} = \frac{1}{n}\sum_{i=1}^n f_i (\z_i^\intercal,\x^\intercal_{i,\hm})(\z_i^\intercal,\x^\intercal_{i,\hm})\transpose,$$
where $f_i = \text{expit}'(\z_i^\intercal\beta + \x_i^\intercal\gamma)$. Here, we assume $f_i$ is known, for $i = 1,\ldots,n$. Later this section, we relax the strong assumption on $f_i$ and assume $f_i$ is unknown. Denote $\bm{\mbox{D}}$ as a diagonal matrix, where $\text{diag}(\bm{\mbox{D}}) = (f_1, \ldots, f_n)$.  Denote $\tilde{\bm{\z}}\transpose = \bm{\z}\transpose\bm{\D}^{1/2}$,  $\tilde{\bm{\mbox{x}}}\transpose = \bm{\x}\transpose\bm{\D}^{1/2}$ and $\tilde{\bm{\x}}_{\hm}^\intercal = \bm{\x}_{\hm}^\intercal\bm{\D}^{1/2}$. Denote the projection matrix as $\bm{\tilde{\bm{\mbox{P}}}_{\hm}} = \tilde{\bm{\x}}_{\hm} (\tilde{\bm{\x}}_{\hm}\transpose\tilde{\bm{\x}}_{\hm})^{-1}\tilde{\bm{\x}}_{\hm}\transpose$.

\subsubsection*{Step 2. Asymptotic consistency}
Now, we want to formally prove the asymptotic consistency of smoothed R-Split estimator under GLM:
\begin{equation}\label{eq:consistency}
    \sqrt{n}a_1\transpose(\tilde{\beta}-\beta) = \eta_n\transpose(a_1)\frac{1}{\sqrt{n}}\sum_{i=1}^n (\z_i\transpose,\x_i\transpose)\transpose \nu_i+ o_p(1),
\end{equation}
\subsubsection*{Step 2 (a). assume $f_i$ is known}
In the first part of the proof, for simplicity, we assume $f_i$ is known. (We will assume $f_i$ is unknown and bound the relevant remainder terms in Step 2  (b)).  $f_i = \text{expit}'(\x_i\transpose\beta+\z_i\transpose\gamma)$ and $f_i$ satisfies Assumption \ref{assumption:differentiability}.  Let $\mbox{I}_{\hm}$ be an index matrix, $\mbox{I}_{\hm} (\z\transpose,\x\transpose)\transpose =(\z\transpose,\x_{\hm}\transpose)\transpose$. Take a subsample $T_2$ of size $n_2$. Assume the subsample is indexed by $\mbox{s} = (\mbox{s}_1,\ldots,\mbox{s}_n)$, where  $\mbox{s}_i = \mathbf{1}_{(i\in T_2)}$. Denote $a_1$ as a random vector, where $||a_1||_2=1$. For the smoothed estimator $\tilde{\beta} = \frac{1}{B}\sum_{i=1}^B \hat{\beta}_{\hm}$,
\begin{align*}
    	\sqrt{n}a_1\transpose(\tilde{\beta} -\beta) &= \mathbb{E}\Big(\sqrt{n}(\hat{\beta}_{\hm} - \beta_{\m_0})|\y,\z,\x\Big),\\
    	&=\frac{1}{\sqrt{n}}\sum_{i=1}^n \mathbb{E}\Big(a_1\transpose\hat{\Sigma}^{-1}_{\hm,\mbox{s}}  \mbox{I}_{\hm}\cdot\mbox{s}_i|\y,\z,\x\Big)(\z_i\transpose,\x_i\transpose)\nu_i \\
	&+ \mathbb{E} \Big(\sqrt{n}(\tilde{\bm{\z}}_{\mbox{s}}\transpose(\bm{\mbox{I}}-\tilde{\bm{\mbox{P}}}_{\hm,\mbox{s}})\tilde{\bm{\z}}_{\s})^{-1}\tilde{\bm{\z}}_{\s}^\intercal(\bm{\mbox{I}}-\tilde{\bm{\mbox{P}}}_{\hm,\s})\tilde{\bm{\x}}_{\s}\beta|\y,\z,\x\Big),\\
	&= \frac{1}{\sqrt{n}}\sum_{i=1}^n \eta_n\transpose(a_1) (\z_i\transpose,\x_i\transpose)\transpose \nu_i+ \mbox{R}_{n1} + \mbox{R}_{n2},
\end{align*}
where 
\begin{align*}
    \mbox{R}_{n1} &= \frac{1}{\sqrt{n}}\sum_{i=1}^n\Big\{\mathbb{E}\Big(a_1\transpose\hat{\Sigma}^{-1}_{\hm,\mbox{s}}  \mbox{I}_{\hm}\cdot\mbox{s}_i|\y,\z,\x\Big) -\eta_n\transpose(a_1) \Big\}(\z_i\transpose,\x_i\transpose)\transpose\nu_i,\\
    \mbox{R}_{n2} &=\mathbb{E} \Big(\sqrt{n}(\tilde{\bm{\z}}_{\mbox{s}}\transpose(\bm{\mbox{I}}-\tilde{\bm{\mbox{P}}}_{\hm,\mbox{s}})\tilde{\bm{\z}}_{\s})^{-1}\tilde{\bm{\z}}_{\s}^\intercal(\bm{\mbox{I}}-\tilde{\bm{\mbox{P}}}_{\hm,\s})\tilde{\bm{\x}}_{\s}\beta|\y,\z,\x\Big).
\end{align*}

We want to show $\mbox{R}_{n1} + \mbox{R}_{n2} = o_p(1)$. Conditioning on $\s_i = 1$,
\begin{align*}
\mathbb{E}\Big(a_1\transpose\hat{\Sigma}^{-1}_{\hm,\mbox{s}}  \mbox{I}_{\hm}\cdot\mbox{s}_i|\y,\z,\x\Big)&= \mathbb{E}\Big(a_1\transpose\hat{\Sigma}^{-1}_{\hm,\mbox{s}}  \mbox{I}_{\hm}\cdot\mbox{s}_i|\y,\z,\x, s_i = 1\Big)\mathbb{P}(\s_i = 1|\y,\z,\x),\\
&= \mathbb{E}\Big(a_1\transpose\hat{\Sigma}^{-1}_{\hm,\mbox{s}}  \mbox{I}_{\hm}|\y,\z,\x,s_i = 1\Big)\cdot r_{\s}.
\end{align*}
Assume there is another subsample indexed by $\tilde{\s} = (\tilde{\s}_1, \ldots, \tilde{\s}_n)$, $\tilde{\s} \perp \s$, where $\tilde{\m}$ is the selected model under $\tilde{\s}$. We can decompose $\mbox{R}_{n1}$ as
\begin{align*}
    \mbox{R}_{n1} & = \underbrace{\Big\{\mbox{r}_{\s}\mathbb{E}\Big(a_1\transpose\hat{\Sigma}^{-1}_{\tilde{M},\tilde{\mbox{s}}}  \mbox{I}_{\tilde{M}}|\y,\z,\x\Big) -\eta_n\transpose(a_1)\Big\}\transpose \frac{1}{\sqrt{n}}\sum_{i=1}^n (\z_i\transpose,\x_i\transpose)\transpose\nu_i}_{\mbox{R}_{n1,1}} ,\\
    &+ \underbrace{\frac{1}{\sqrt{n}} \sum_{i=1}^n \Big\{\mathbb{E}\Big(a_1\transpose\hat{\Sigma}^{-1}_{\hm,\mbox{s}}  \mbox{I}_{\hm}\cdot \s_i|\y,\z,\x\Big) - \mathbb{E}\Big(a_1\transpose\hat{\Sigma}^{-1}_{\tilde{M},\tilde{\mbox{s}}}  \mbox{I}_{\tilde{M}}\cdot \s_i|\y,\z,\x\Big)\Big\}\transpose(\z\transpose,\x\transpose)\transpose\nu_i}_{\mbox{R}_{n1,2}}.
\end{align*}
By Assumption \ref{assumption:overfitting}, $\mbox{R}_{n1,2} = o_p(1)$. By H$\ddot{\text{o}}$lder's inequality, Assumption \ref{assumption:random-vector} and Lemma \ref{lemma1}, 
\begin{align*}
    \mbox{R}_{n1,1} \leq \Big|\Big|\mbox{r}_{\s}\mathbb{E}\Big(a_1\transpose(\hat{\Sigma}_{\tilde{\m},\tilde{\s}})^{-1}\mbox{I}_{\tilde{\m}} | \y,\z,\x\Big) -\eta\transpose_n(a_1) \Big|\Big|_1 \cdot \Big|\Big|(\bm{\z}\transpose, \bm{\x}\transpose)\transpose\nu\sqrt{n}\Big|\Big|_{\infty} =  o_p(1).
\end{align*}
Therefore, $\mbox{R}_{n1} + \mbox{R}_{n2} = o_p(1)$.

\subsubsection*{Step 2 (b). assume $f_i$ is unknown}
Next, we assume $f_i$ is unknown and bound the remainder terms related to $f_i$. Denote $f_i = \text{expit}'(\z_i\transpose\beta + \x_i\transpose\gamma)$ and $\hat{f}_i = \text{expit}'(\z_i\transpose\hat{\beta} + \x_i\transpose\hat{\gamma})$. Denote  the sample Hessian matrix under true $f$ as $\hat{\Sigma}_{\hm,f}$ and under estimated $\hat{f}$ as $\hat{\Sigma}_{\hm,\hat{f}}$. Similarly, we can define $\tilde{\bm{\z}}_{f}$ and $\tilde{\bm{\z}}_{\hat{f}}$, $\tilde{\bm{\x}}_{\hm,f}$ and $\tilde{\bm{\x}}_{\hm,\hat{f}}$, $\tilde{\bm{\mbox{P}}}_{\hm,f}$ and $\tilde{\bm{\mbox{P}}}_{\hm,\hat{f}}$. Denote $\nu_i = \big(\y_i-\text{expit}(\z_i\transpose\beta + \x_i\transpose\gamma)\big)$ and $\hat{\nu}_i = \big(\y_i -\text{expit}(\z_i\transpose\hat{\beta} + \x_i\transpose\hat{\gamma})\big)$. 
\begin{align*}
	\sqrt{n}(\hat{\beta}_{\texttt{GLM}} -\beta)&= \mbox{I}_{\z}\hat{\Sigma}^{-1}_{\hm,f}\cdot \frac{1}{\sqrt{n}}\sum_{i=1}^n (\z_i\transpose,\x_{i,\hm}\transpose)\transpose\nu_i + (\tilde{\bm{\z}}_{\hf}\transpose(\bm{\mbox{I}}-\tilde{\bm{\mbox{P}}}_{\hm,\hf})\tilde{\bm{\z}}_{\hf}/n)^{-1}\tilde{\bm{\z}}_{\hf}\transpose(\bm{\mbox{I}}-\tilde{\bm{\mbox{P}}}_{\hm,\hf})\tilde{\bm{\x}}_{\hf}\beta/\sqrt{n} + \mbox{R}'_n,\\
\mbox{R}'_{n}&= \mbox{R}'_{n1} + \mbox{R}'_{n2},\\ &=\Big(\mbox{I}_{\z}\hat{\Sigma}^{-1}_{\hm,\hf}\cdot \frac{1}{\sqrt{n}}\sum_{i=1}^n (\z_i\transpose,\x_{i,\hm}\transpose)\transpose\nu_i - \mbox{I}_{\z}\hat{\Sigma}^{-1}_{\hm,f}\cdot \frac{1}{\sqrt{n}}\sum_{i=1}^n (\z_i\transpose,\x_{i,\hm}\transpose)\transpose\nu_i\Big) \\
&+ \Big(\mbox{I}_{\z}\hat{\Sigma}^{-1}_{\hm,\hf}\cdot \frac{1}{\sqrt{n}}\sum_{i=1}^n (\z_i\transpose,\x_{i,\hm}\transpose)\transpose\hat{\nu}_i - \mbox{I}_{\z}\hat{\Sigma}^{-1}_{\hm,\hf}\cdot \frac{1}{\sqrt{n}}\sum_{i=1}^n (\z_i\transpose,\x_{i,\hm}\transpose)\transpose\nu_i\Big).
\end{align*}

We want to show $\mbox{R}_n'$ is $O_p\Big(\frac{|\m_0|\log(p)^{3/2+\delta}}{n}\Big)$. For simplicity, denote $\w_{i,\hm} =(\z_i\transpose, \x_{i,\hm}\transpose)$.
\begin{align*}
 \mbox{R}'_{n1} &=  a_1\transpose\Big( \hat{\Sigma}^{-1}_{\hm,\hf} -\hat{\Sigma}^{-1}_{\hm,f} \Big)\Big(\frac{1}{\sqrt{n}}\sum_{i=1}^n \w_{i,\hm}\transpose\nu_i\Big),\\
 &= a_1\transpose\hat{\Sigma}^{-1}_{\hm,\hf} \Big( \hat{\Sigma}_{\hm,f} -\hat{\Sigma}_{\hm,\hf} \Big)\hat{\Sigma}^{-1}_{\hm,f}\Big(\frac{1}{\sqrt{n}}\sum_{i=1}^n \w_{i,\hm}\transpose\nu_i\Big),\\
 &= a_1\transpose\hat{\Sigma}^{-1}_{\hm,\hf}\Big\{ \Big(\frac{1}{n}\sum_{i=1}^n f_i\w_{i,\hm}\w_{i,\hm}\transpose\Big) - \Big(\frac{1}{n}\sum_{i=1}^n \hf_i\w_{i,\hm}\w_{i,\hm}\transpose\Big) \Big\}\hat{\Sigma}^{-1}_{\hm,f} \Big(\frac{1}{\sqrt{n}}\sum_{i=1}^n \w_{i,\hm}\transpose\nu_i\Big) ,\\
 &\leq   \max_i\Big| \Big(\frac{1}{n}\sum_{i=1}^n f_i\w_{i,\hm}\w_{i,\hm}\transpose\Big) - \Big(\frac{1}{n}\sum_{i=1}^n  \hf_i\w_{i,\hm}\w_{i,\hm}\transpose\Big) \Big|\cdot  \Big|\Big|a_1\transpose\hat{\Sigma}^{-1}_{\hm,\hf}\Big|\Big|_2\Big|\Big|\hat{\Sigma}^{-1}_{\hm,f}\frac{1}{\sqrt{n}}\sum_{i=1}^n \w_{i,\hm}\transpose\nu_i \Big|\Big|_2 .
\end{align*}
To bound 
$\max_i\Big| \frac{1}{n}\sum_{i=1}^n (\hf_i - f_i)\w_{i,\hm}\w_{i,\hm}\transpose\Big|$, we first work with $(\hf_i - f_i)$.
\begin{align*}
    \hf_i - f_i &= \text{expit}'(\z_i\transpose\hat{\beta} + \x_i\transpose\hat{\gamma}) - \text{expit}'(\z_i\transpose\beta + \x_i\transpose\gamma),\\
    &= \text{expit}(\z_i\transpose\hat{\beta} + \x_i\hat{\gamma})\big(1-\text{expit}(\z_i\transpose\hat{\beta}+\x_i\hat{\gamma})\big) - \text{expit}(\z_i\transpose\beta + \x_i\transpose\gamma) (1-\text{expit}(\z_i\transpose\beta + \x_i\transpose\gamma)),\\
    &= \Big(\text{expit}(\z_i\transpose\hat{\beta} + \x_i\hat{\gamma}) - \text{expit}(\z_i\transpose\beta + \x_i\transpose\gamma)\Big) + \Big(\text{expit}^2(\z_i\transpose\beta + \x_i\gamma) - \text{expit}^2(\z_i\transpose\hat{\beta} + \x_i\transpose\hat{\gamma})\Big),\\
    &= \Big(\text{expit}(\z_i\transpose\hat{\beta} + \x_i\hat{\gamma}) - \text{expit}(\z_i\transpose\beta + \x_i\transpose\gamma)\Big) \\
    &+ \Big(\text{expit}(\z_i\transpose\beta + \x_i\gamma) +
    \text{expit}(\z_i\transpose\hat{\beta} + \x_i\transpose\hat{\gamma})\Big)\Big(\text{expit}(\z_i\transpose\beta + \x_i\gamma) - \text{expit}(\z_i\transpose\hat{\beta} + \x_i\transpose\hat{\gamma})\Big).
\end{align*}
Thus
   $ \max_i\Big|\frac{1}{n}\sum_{i=1}^n (\hf_i-f_i)\w_{i,\hm}\w_{i,\hm}\transpose\Big| =  C \cdot O_p(\frac{|\m_0|\log(p)^{3/2+\delta}}{n})$ by Assumption \ref{assumption:data} and Assumption \ref{assumption: glm-lasso-rate}. The last two $l_2$ norms in $\mbox{R}'_{n1}$ can be bounded by Lemma \ref{lemma1}, with convergence rate $O_p (|\hm|^{1/2}\sqrt{\log p})$. In sum,
   \begin{align*}
       \mbox{R}'_{n1} &= C \cdot O_p(\frac{|\m_0|\log(p)^{3/2+\delta}}{n})\cdot  O_p (|\hm|^{1/2}\sqrt{\log p}).
   \end{align*}

Next, we bound the remainder term $\mbox{R}
'_{n2}$.
\begin{align*}
    \mbox{R}'_{n2} &= \Big(a_1\transpose\hat{\Sigma}^{-1}_{\hm,\hf}\cdot \frac{1}{\sqrt{n}}\sum_{i=1}^n \w_{i,\hm}\transpose\hat{\nu}_i - a_1\transpose\hat{\Sigma}^{-1}_{\hm,\hf}\cdot \frac{1}{\sqrt{n}}\sum_{i=1}^n \w_{i,\hm}\transpose\nu_i\Big),\\
    &= \Big(a_1\transpose\hat{\Sigma}^{-1}_{\hm,\hf}\cdot \frac{1}{\sqrt{n}}\sum_{i=1}^n \w_{i,\hm}\transpose(\hat{\nu}_i -\nu_i)\Big),\\
    &= \Big(a_1\transpose\hat{\Sigma}^{-1}_{\hm,\hf}\cdot \frac{1}{\sqrt{n}}\sum_{i=1}^n \w_{i,\hm}\transpose\big(\text{expit}(\z_i\transpose\hat{\beta} + \x_i\transpose\hat{\gamma}) -\text{expit}(\z_i\transpose\beta + \x_i\transpose\gamma)\big)\Big),\\
    &\leq \sqrt{n}\cdot \lambda^{-1}_{\min}(a_1\transpose\hat{\Sigma}_{\hm,\hf})\cdot \sqrt{\mathbb{E} \Big[\big(\w_{i,\hm}\transpose\big)^2\Big]\mathbb{E}\Big[\Big(\text{expit}(\z_i\transpose\hat{\beta} + \x_i\transpose\hat{\gamma}) -\text{expit}(\z_i\transpose\beta + \x_i\transpose\gamma)\Big)^2\Big]}, \\
    &=O_p\Big(\frac{|\m_0|\log(p)^{3/2+\delta}}{n}\Big)
\end{align*}
by H$\ddot{\text{o}}$lder's inequality, Assumption \ref{assumption:data} and Assumption \ref{assumption: glm-lasso-rate}. In sum, $\mbox{R}'_n = \mbox{R}'_{n1} + \mbox{R}'_{n2} = O_p\Big(\frac{|\m_0|\log(p)^{3/2+\delta}}{n}\Big)$. Combining Step 2 (a) and Step 2 (b), we prove the consistency result in Equation (\ref{eq:consistency}). 

\subsubsection*{Step 3. Asymptotic normality}

Now let $\tilde{\sigma}_{n}(a_1) = \sigma_{\nu}(\eta_n(a_1)\transpose\hat{\Sigma}_n\eta_n(a_1))^{1/2}$, where $\hat{\Sigma}_n = \frac{1}{n}\sum_{i=1}^n f_i(\z_i\transpose, \x_i\transpose)(\z_i\transpose, \x_i\transpose)\transpose$. Under Assumptions \ref{assumption:data}, Assumptions \ref{assumption:differentiability} and Assumption \ref{assumption: glm-lasso-rate} and the asymptotic consistency result in Equation (\ref{eq:consistency}), we have
\begin{equation}\label{eq:normality}
    \tilde{\sigma}^{-1}_{n}(a_1) \sqrt{n}a_1\transpose(\tilde{\beta} - \beta) \leadsto N(0,1).
\end{equation}
\end{proof}

\subsubsection{R-Split's bootstrap consistency under GLM}\label{appendix:bootstrap-consistency}

In this section, we want to prove R-Split's bootstrap consistency by showing:
	\begin{equation}\label{eq:bootstrap-consistency}
	   a_1\transpose
	  (\tilde{\beta}^*-\tilde{\beta}) =  \eta_n\transpose(a_1) \cdot \frac{1}{n} \sum_{i=1}^n (\z_i\transpose, \x_i\transpose)\transpose \nu_i + o_p(1).
	\end{equation}
By the construction of residual bootstrap, we have 
	\begin{align*}
	    a_1\transpose( \tilde{\beta}^*-\tilde{\beta}) =  \tilde{\eta}\transpose_n(a_1) \cdot \frac{1}{n} \sum_{i=1}^n (\z_i\transpose, \x_i\transpose)\transpose \nu_i^*,
	\end{align*}
where $\nu_i^* = u_i\hat{\nu}_i$. Following a direct expansion of the bootstrap approximation: 
	\begin{align*}
	    \tilde{\eta}\transpose_n(a_1) \cdot \frac{1}{n} \sum_{i=1}^n (\z_i\transpose, \x_i\transpose)\transpose \nu_i^* = {\eta}\transpose_n(a_1)\cdot \frac{1}{n} \sum_{i=1}^n (\z_i\transpose, \x_i\transpose)\transpose  u_i\nu_i + r_{n1} + r_{n2},
	\end{align*}
	where the remainder terms $ r_{n1}  = \big(\tilde{\eta}_n(a_1) - \eta_n(a_1)\big)\transpose \cdot \frac{1}{n} \sum_{i=1}^n  (\z_i\transpose, \x_i\transpose)\transpose u_i\nu_i $ and $r_{n2} = \tilde{\eta}_n\transpose(a_1)\cdot \frac{1}{n} \sum_{i=1}^n (\z_i\transpose, \x_i\transpose)\transpose u_i (\hat{\nu}_i - \nu_i)$.
	Since $u_i$'s are i.i.d. random variable with mean 0 and variance 1, the leading term has the same distribution as $\tilde{\beta} - \beta$. Now we define the maximal eigenvalue of matrix $ \frac{1}{n} \sum_{i=1}^n  (\z_i\transpose, \x_i\transpose)\transpose u_i(\hat{\nu}_i - \nu_i)$ to be $\lambda_{\max}$, and we assume this quantity is bounded away from infinity. To prove bootstrap consistency, it is suffice to show that the remainder terms vanish at root-$n$ rates. Under the same Assumptions for the previous consistency proof, we immediately have $r_{n1} = o_p(1/\sqrt{n})$. As for the second remainder term, under Assumption \ref{assumption:data} and Assumption \ref{assumption: glm-lasso-rate}, we have the following bound  $r_{n2} \lesssim_p ||  {\eta}_n||_2 \cdot \lambda_{\max} \cdot || \hat{\nu} - \nu||_2 = O_p\Big(\frac{|\m_0|\log(p)^{3/2+\delta}}{n}\Big)$. 
	
	Following the bootstrap consistency result, we can show $
    \tilde{\sigma}^{-1}_{n}(a_1) \sqrt{n}a_1\transpose(\tilde{\beta}^* - \tilde{\beta}) \leadsto N(0,1)$ in probability, where $\tilde{\sigma}_n(a_1) = \sigma_{\nu}\big(\eta\transpose_n(a_1)\hat{\Sigma}_n\eta_n(a_1)\big)^{1/2}$ and $\hat{\Sigma}_n = \frac{1}{n}\sum_{i=1}^n f_i(\z_i\transpose, \x_i\transpose)(\z_i\transpose, \x_i\transpose)\transpose$.

	\subsection{Proof of Theorem 1}\label{appendix:theorem1-proof}

By C.1 and C.2, we have under Assumptions 1-7, for any $a\in \mathbb{R}^{p_1}$ 
\begin{equation}\label{basic}
    \tilde{\sigma}^{-1}_{n}(a)\sqrt{n}a_1\transpose(\tilde{\beta}-\beta)\to N(0,1); \hspace{0.2cm} \tilde{\sigma}^{-1}_{n}(a)\sqrt{n}a_1\transpose(\tilde{\beta}^*-\tilde{\beta})\to N(0,1)\textit{ in probability,} 
\end{equation}
 
Because under Assumption 4, by definition, we can construct a series of $\eta_n(a)$ satisfying additive property; i.e. $\eta_{n}(k_1c+k_2d)=k_1\eta_{n}(c)+k_2\eta_{n}(d)$ for any vectors c and d and constant $k_1$ and $k_2$. To be specific, let $\eta_n(e_i)$ denote the random vectors satisfying Assumption 4 for the base vector $e_i\in \mathbb{R}^{p_1}$ for $i=1,\dots,p_1$. Then, for any vector $a=\sum_{i=1}^{i=p_1}k_ie_i$, letting $\eta_n(a)=k_i\eta_n(e_i)$ naturally satisfies assumptions 4 as illustrated below, 
\begin{align*}
   \Big|\Big|\mbox{r}_{\s}\mathbb{E}\Big(a\transpose\hat{\Sigma}^{-1}_{\hm,\s}\mbox{I}_{\hm}|\y,\z,\x\Big) - \eta\transpose_n(a_1)\Big|\Big|_1 =  \Big|\Big|\sum_{i=1}^{i=p_1}k_i[\mbox{r}_{\s}\mathbb{E}\Big(e_i\transpose\hat{\Sigma}^{-1}_{\hm,\s}\mbox{I}_{\hm}|\y,\z,\x\Big) - \eta\transpose_n(e_i)]\Big|\Big|_1=o_p(1/\sqrt{\log p}), 
\end{align*}
because by definition, $\Big|\Big|\mbox{r}_{\s}\mathbb{E}\Big(e_i\transpose\hat{\Sigma}^{-1}_{\hm,\s}\mbox{I}_{\hm}|\y,\z,\x\Big) - \eta\transpose_n(e_i)\Big|\Big|_1=o_p(1/\sqrt{\log p})$ for $i=1,\dots,p_1$. Let $(\xi_{1,n},\dots,\xi_{p_1,n})\sim N(0,\tilde{\Sigma}_n)$ be a multivariate normal distribution where $\tilde{\Sigma}_{n}=\sigma_{\nu}(L_n\hat{\Sigma}_nL_n^T)$ and $L_n=(\eta_n(e_1),\dots,\eta_n(e_{p_1}))^T$. Then, we have $\tilde{\sigma}_{n}(a)=a\transpose\tilde{\Sigma}_n a$. By Eq. \ref{basic} and Assumption 8, because $p_1$ is fixed, we have

\begin{equation}\label{advance}
    \sqrt{n}\tilde{\Sigma}_{n}^{-1/2}(\tilde{\beta}-\beta)\to N(0,I_p); \hspace{0.2cm} \sqrt{n}\tilde{\Sigma}_{n}^{-1/2}(\tilde{\beta}^*-\tilde{\beta})\to N(0,I_p)\textit{ in probability,} 
\end{equation}

To prove Theorem 1, we prove the following results: 
	\begin{align*}
	    & \sup_{c\in\mathbb{R}} \left| \mathbb{P}\left( \sqrt{n}\left(\max_{j\in [p_1]}\tilde{\beta}_j - {\beta}_{\max}\right) \leq c \right) - \mathbb{P}\left( \max_{j\in H}\sqrt{n}( \tilde{\beta}_j - \beta_j )\leq c \right)\right|  = o(1),\\
	    & \sup_{c\in\mathbb{R}} \left| \mathbb{P}^*\left( \sqrt{n}\left(\max_{j\in [p_1]}( \tilde{\beta}^*_j + \tilde{c}_j(r )) - \tilde{\beta}_{\max}\right) \leq c \right) - \mathbb{P}\left( \max_{j\in H}\sqrt{n}( \tilde{\beta}_j - \beta_j )\leq c \right)\right|  = o_p(1).
	\end{align*}
	
\textit{First}, for the R-Split estimate, we have the following arguments: 
	\begin{align*}
	\sqrt{n}\left(\max_{j\in [p_1]}\tilde{\beta}_j - {\beta}_{\max}\right) =& \max_{j\in [p_1]}\left(\sqrt{n}\tilde{\beta}_j + \sqrt{n}({\beta}_{j}-{\beta}_{j}) - \sqrt{n}{\beta}_{\max}\right) \\
	=& \max_{j\in [p_1]}\left(\sqrt{n} (\tilde{\beta}_j -{\beta}_{j})+\sqrt{n}({\beta}_{j} - {\beta}_{\max}) \right),
	\end{align*}
	then 
	\begin{align*}
	& \left| \mathbb{P}\left( \sqrt{n}\left(\max_{j\in [p_1]}\tilde{\beta}_j - {\beta}_{\max}\right) \leq c \right) - \mathbb{P}\left( \max_{j\in H}\sqrt{n}( \tilde{\beta}_j - \beta_j )\leq c \right)\right| \\
	= &\left| 	\mathbb{P}\left(\max_{j\in [p_1]}\left(\sqrt{n} (\tilde{\beta}_j -{\beta}_{j}) + \sqrt{n}(\beta_j - {\beta}_{\max}) \right)\leq c \right) - \mathbb{P}\left( \max_{j\in H}\sqrt{n}( \tilde{\beta}_j - \beta_j )\leq c \right)\right| \\
	=&\Big|  \mathbb{P}\left(\sqrt{n} (\tilde{\beta}_j -{\beta}_{j}) \leq c, \text{ for } j\in H;\  \sqrt{n} (\tilde{\beta}_j -{\beta}_{j}) + \sqrt{n}(\beta_j - {\beta}_{\max}) \leq c,\text{ for } j\notin H \right)\\
	&- \mathbb{P}\left( \max_{j\in H}\sqrt{n}( \tilde{\beta}_j - \beta_j )\leq c \right) \Big|\\
	\le & \mathbb{P}\left(  \sqrt{n} (\tilde{\beta}_j -{\beta}_{j}) + \sqrt{n}(\beta_j - {\beta}_{\max}) > c,\text{ any } j\notin H \right)\\
	=&  1-\mathbb{P}\left(  \sqrt{n} (\tilde{\beta}_j -{\beta}_{j}) + \sqrt{n}(\beta_j - {\beta}_{\max}) \leq c,\text{ for } j\notin H \right).
	\end{align*}
	Therefore, given any fixed $c_0\in \mathbb{R}$ and for any $c>c_0$, we have 
	\begin{align*}
	& \left| \mathbb{P}\left( \sqrt{n}\left(\max_{j\in [p_1]}\tilde{\beta}_j - {\beta}_{\max}\right) \leq c \right) - \mathbb{P}\left( \max_{j\in H}\sqrt{n}( \tilde{\beta}_j - \beta_j )\leq c \right)\right| \\ 
	\leq  &  1-\mathbb{P}\left(  \sqrt{n} (\tilde{\beta}_j -{\beta}_{j}) + \sqrt{n}(\beta_j - {\beta}_{\max}) \leq c_0,\text{ for } j\notin H \right) .
	\end{align*}
	By taking supreme on both side, we obtain 
	\begin{align*}
	& \sup_{c>c_0} \left| \mathbb{P}\left( \sqrt{n}\left(\max_{j\in [p_1]}\tilde{\beta}_j - {\beta}_{\max}\right) \leq c \right) - \mathbb{P}\left( \max_{j\in H}\sqrt{n}( \tilde{\beta}_j - \beta_j )\leq c \right)\right| \\ 
	\leq  &  1-\mathbb{P}\left(  \sqrt{n} (\tilde{\beta}_j -{\beta}_{j}) \leq c_0 + \sqrt{n}({\beta}_{\max} -\beta_j  ) ,\text{ for } j\notin H \right).
	\end{align*}
	By Assumptions 8 and 9 and Eq. \ref{advance}, the quantity on the right hand side is upper bounded by $\max_{j\in [p_1]}\beta_j-\max_{j\notin H}\beta_j$ and the upper bound of $\tilde{\Sigma}_{n;i,i}$ over $i\in [p_1]$-- the diagnol of $\tilde{\Sigma}_{n}$ as follows
    \begin{align*}
    & 1-\mathbb{P}\left(  \sqrt{n} (\tilde{\beta}_j -{\beta}_{j}) \leq c_0 + \sqrt{n}({\beta}_{\max} -\beta_j  ) ,\text{ for } j\notin H \right)\\
     \lesssim & \mathbb{P}\left( \max_{j\notin H} \xi_{j,n} \geq   c_0 + \sqrt{n}(\max_{j\in [p_1]}\beta_j-\max_{j\notin H}\beta_j )\right) \\
    \leq & \mathbb{P}\left( \max_{j\in [p_1] } \xi_{j,n} \geq   c_0 + \sqrt{n}(\max_{j\in [p_1]}\beta_j-\max_{j\notin H}\beta_j )\right) \\
    \leq &\sum_{i=1}^{i=p_1}\exp\left(- (c_0+\sqrt{n}(\max_{j\in [p_1]}\beta_j-\max_{j\notin H}\beta_j))^2/2\tilde{\Sigma}_{n;i,i} ) \right)\\
    \leq &  \exp\left( \log p_1 - \sqrt{n}c_0(\max_{j\in [p_1]}\beta_j-\max_{j\notin H}\beta_j )/U \right) = o(1).
    \end{align*}
	where $a_n\lesssim b_n$ indicates that $\lim\sup_{n\to\infty} a_n\le b_n$. 
	
	Since, by Assumption 9, $\limsup_{c_0\rightarrow -\infty} \mathbb{P}( \max_{j\in H}\sqrt{n} (\tilde{\beta}_j -{\beta}_{j}) \leq c_0  ) =0$, we have the following statement: 
	\begin{align}\label{eq:thm1-key2}
	   \sup_{c\in\mathbb{R}} \left| \mathbb{P}\left( \sqrt{n}\left(\max_{j\in [p_1]}\tilde{\beta}_j - {\beta}_{\max}\right) \leq c \right) - \mathbb{P}\left( \max_{j\in H}\sqrt{n}( \tilde{\beta}_j - \beta_j )\leq c \right)\right|  = o(1).
	\end{align}
	
	\textit{Second}, since $\tilde{c}_j(r) =(1-n^{r-0.5})(\tilde{\beta}_{\max}-\tilde{\beta}_{j})$ the modified bootstrap estimate satisfies 
	\begin{align*}
    \sqrt{n}\left(\max_{j\in [p_1]}( \tilde{\beta}_j^* + \tilde{c}_j(r )) - \tilde{\beta}_{\max}\right) =& \max_{j\in [p_1]}\left(\sqrt{n}\tilde{\beta}_j^* + (\sqrt{n}-n^{r})(\tilde{\beta}_{\max}-\tilde{\beta}_{j}) - \sqrt{n}\tilde{\beta}_{\max}\right) \\
    =& \max_{j\in [p_1]}\left(\sqrt{n} (\tilde{\beta}_j^* -\tilde{\beta}_{j})+n^{r}(\tilde{\beta}_{j} - \tilde{\beta}_{\max}) \right).
	\end{align*}
	Therefore,	for $c\in\mathbb{R}$, the distribution of the modified bootstrap estimation has
	\begin{align*}
	 &\mathbb{P}^*\left( \sqrt{n}\left(\max_{j\in [p_1]}( \tilde{\beta}_j^* + \tilde{c}_j(r )) - \tilde{\beta}_{\max}\right) \leq c \right) \\
	= &	\mathbb{P}^*\left(\max_{j\in [p_1]}\left(\sqrt{n} (\tilde{\beta}_j^* -\tilde{\beta}_{j})+n^{r}(\tilde{\beta}_{j} -\beta_j+\beta_j - \beta_{\max} + \beta_{\max} - \tilde{\beta}_{\max}) \right)\leq c \right) \\
	=& \mathbb{P}^*\big(\sqrt{n} (\tilde{\beta}_j^* -\tilde{\beta}_{j}) \leq c-n^{r}(\tilde{\beta}_{j} -\beta_j + \beta_{\max} - \tilde{\beta}_{\max}), \text{ for }j\in H, \\
	&\sqrt{n} (\tilde{\beta}_j^* -\tilde{\beta}_{j})\leq c-n^{r}(\tilde{\beta}_{j} -\beta_j + \beta_{\max} - \tilde{\beta}_{\max}) + n^{r}(\beta_{\max}-\beta_j), \text{ for }j\notin H \big).
	\end{align*}
	Similar to the first part, given any fixed $c_0\in \mathbb{R}$ and for any $c>c_0$, we have 
	\begin{align}
	   &\nonumber \sup_{c>c_0} \Big| \mathbb{P}^*\left(\sqrt{n}(\max_{j\in [p_1]}( \tilde{\beta}_j^* + \tilde{c}_j(r )) - \tilde{\beta}_{\max}) \leq c \right) \\
	   &\nonumber -\mathbb{P}^* \left(\sqrt{n} (\tilde{\beta}_j^* -\tilde{\beta}_{j}) \leq c-n^{r}(\tilde{\beta}_{j} -\beta_j + \beta_{\max} - \tilde{\beta}_{\max}), \text{ for }j\in H\right) \Big| \\
	   \leq & 1 - \mathbb{P}^*\left(\sqrt{n} (\tilde{\beta}_j^* -\tilde{\beta}_{j})\leq c_0-n^{r}(\tilde{\beta}_{j} -\beta_j + \beta_{\max} - \tilde{\beta}_{\max}) + n^{r}(\beta_{\max}-\beta_j), \text{ for }j\notin H\right).\label{eq:thm-key1}
	\end{align}
    For the right hand side of \eqref{eq:thm-key1}, recall that under Assumption 8, we have $\max_{j\in [p_1]} n^{r}|\tilde{\beta}_{j} -\beta_j + \beta_{\max} - \tilde{\beta}_{\max}| = o_p(1)$. By Assumption 8, we have 
    \begin{align*}
        \max_{j\in [p_1]} (n^{r}|\tilde{\beta}_{j} -\beta_j + \beta_{\max} - \tilde{\beta}_{\max}|/\sqrt{\hat{\Sigma}_{n;j,j}})
        \leq \max_{j\in [p_1]}( n^{r}|\tilde{\beta}_{j} -\beta_j + \beta_{\max} - \tilde{\beta}_{\max}|/\sqrt{L}) = o_p(1)
    \end{align*}
    Therefore, by anti-concentration inequality, we have
	\begin{align*}
	      &\mathbb{P}^*\left(\sqrt{n} (\tilde{\beta}_j^* -\tilde{\beta}_{j})\leq c-n^{r}(\tilde{\beta}_{j} -\beta_j + \beta_{\max} - \tilde{\beta}_{\max}) + n^{r}(\beta_{\max}-\beta_j), \text{ for }j\notin H\right) \\
	      &-\mathbb{P}\left(  \sqrt{n} (\tilde{\beta}_j -{\beta}_{j}) \leq c + n^r({\beta}_{\max} -\beta_j  ) ,\text{ for } j\notin H \right) = o_p(1),  
	\end{align*}
	uniformly in $c>c_0$. By Assumptions 8 and 9, we can show that
	\begin{align*}
	    1-\mathbb{P}\left(  \sqrt{n} (\tilde{\beta}_j -{\beta}_{j}) \leq c + n^r({\beta}_{\max} -\beta_j  ) ,\text{ for } j\notin H \right)=o(1),
	\end{align*}
	uniformly in $c>c_0$. Therefore, the right hand side of \eqref{eq:thm-key1} converges to 0 in probability. For the left hand side of \eqref{eq:thm-key1}, again by Assumption 8 and anti-concentration inequality, we can also show that 
	\begin{align*}
	  &\nonumber \sup_{c>c_0} \left| \mathbb{P}^* \left(\sqrt{n} (\tilde{\beta}_j^* -\tilde{\beta}_{j}) \leq c-n^{r}(\tilde{\beta}_{j} -\beta_j + \beta_{\max} - \tilde{\beta}_{\max}), \text{ for }j\in H\right) -  \mathbb{P}\left( \max_{j\in H}\sqrt{n}( \tilde{\beta}_j - \beta_j )\leq c \right)\right|\\ 
	  =& o_p(1). 
	\end{align*}
	By the similar argument we made in the first part, we have shown that 
	\begin{align*}
	   \sup_{c\in\mathbb{R}} \left| \mathbb{P}^*\left( \sqrt{n}(\max_{j\in [p_1]}( \tilde{\beta}^*_j + \tilde{c}_j(r )) - \tilde{\beta}_{\max}) \leq c \right) - \mathbb{P}\left( \max_{j\in H}\sqrt{n}( \tilde{\beta}_j - \beta_j )\leq c \right)\right|  = o_p(1).
	\end{align*}
	This result, together with \eqref{eq:thm1-key2} finishes the proof of Theorem 1.

\begin{coro}[Selected subgroup with the maximal treatment effect]\label{corollary:selected-subgroup}
	Under Assumptions 1-9, we have 
		$$\sup_{c\in\mathbb{R}}|\mathbb{P}(\sqrt{n}(\hat{b}_{\max}-\beta_{\hat{s}})\le c)-\mathbb{P}^*(\sqrt{n}(\hat{b}^*_{\mathrm{modified};\max}-\tilde{\beta}_{\max})\le c)|=o_p(1).$$
\end{coro}

\begin{proof}
Let M denote the event $\beta_s<\beta_{\max}$ and $s_0$ denote the one of the best subgroup; i.e. $\beta_{s_0}=\beta_{\max}$. We have
\begin{equation}
    P(M)\leq P(\tilde{\beta}_{s_0}<\max_{i\notin H}\tilde{\beta}_i)\leq \sum_{i\notin H}P(\tilde{\beta}_{s_0}<\tilde{\beta}_i).
\end{equation}
Because for any $i\notin H$, by Assumptions 8 and 9, we have  $$P(\tilde{\beta}_{s_0}<\tilde{\beta}_i)=P(\sqrt{n}(\tilde{\beta}_{s_0}-\beta_{s_0}-\tilde{\beta}_i+\beta_i)<\sqrt{n}(\beta_i-\beta_{s_0}))\to 0$$
as $\sqrt{n}(\beta_i-\beta_{s_0})\to -\infty$. We prove the corollary. 
\end{proof}

\section{Simulation results: Inference on $\beta_{\hat{s}}$}
In this section, we provide additioinal simulatioin results when the inference target is $\beta_{\hat{s}}$, where $\hat{s}= \arg\max_{j\in[p_1]}\hat{\beta}_j$. $\beta_{\hat{s}}$ denotes the true treatment effect of the selected subgroup $\hat{s}$. The simulation results are summarized in Table \ref{table:beta_s}.
\begin{table}[h!]
\caption{ Simulation results (heterogeneous case)}\label{table:beta_s}

\begin{adjustbox}{width=.7\textwidth,center}
\small{\begin{tabular}{cccccc}
     \hline\hline
           & &\multicolumn{3}{c}{  $\beta=(0,\dots,0,1)\in \mathbb{R}^{p_1}$ (heterogeneity)}& \\
           \cline{2-5}
           &&\multicolumn{3}{c}{Logistic Regression ($\bm{p_2=150}$)} &  \\
             & &  Boot-Calibrated & No adjustment & Simultaneous  &  \\
       \cline{2-5} 
    $p_1=4$ &Cover  & 0.95(0.02) & 0.87(0.02)& 0.99(0.01) &
    \\  
    
    & $\sqrt{n}$Length & 9.39(0.02) & 8.02(0.06)& 14.8(0.03) &
      \\
    
    & $\sqrt{n}$Bias &-3.74(3.37)  &5.40(4.09)&--- &
    \\[0.15cm]

    $p_1=10$ &Cover & 0.92(0.01) & 0.85(0.02)& 0.98(0.01) & \\
    
     &$\sqrt{n}$Length & 10.8(0.05)  &9.01(0.02) & 16.7(0.02) &\\ 
     
     & $\sqrt{n}$Bias &-4.77(4.03)  &6.06(5.49) &--- &\\[0.15cm]
      \cline{2-5}

     & & \multicolumn{3}{c}{Repeated Sample Splitting ($\bm{p_2=150}$)} &   \\
     &  & Boot-Calibrated & No adjustment & Simultaneous &  \\  
      \cline{2-5}
     $p_1=4$ &Cover & 0.95(0.01) & 0.93(0.02) &0.99(0.01)&  
     \\
      
      &$\sqrt{n}$ Length & 3.63(0.07) &2.22(0.05) &5.26(0.06)  &\\
      
       & $\sqrt{n}$Bias & 0.15(0.32) &0.19(0.27) &--- &
       \\[0.15cm]
       
       $p_1=10$ &Cover & 0.94(0.02) &0.92(0.01) &0.99(0.01)& \\
       
       &$\sqrt{n}$Length &3.66(0.05) &2.61(0.05) &6.63(0.06)&  \\
       
        & $\sqrt{n}$Bias & 0.30(0.42) &0.35(0.32) &--- & \\
        
       \cline{2-5} 
       
& & \multicolumn{3}{c}{Repeated Sample Splitting ($\bm{p_2=500}$)} &   \\
     &  & Boot-Calibrated & No adjustment & Simultaneous &  \\  
     \cline{2-5}
   $p_1=4$ &Cover &0.94(0.02) & 0.92(0.03) & 0.99(0.00) &\\ 
&$\sqrt{n}$Length &  4.45(0.05) &  2.25(0.05)& 6.10(0.06) & \\
   & $\sqrt{n}$Bias & -0.72(0.91) & 1.25(1.21)& --- &
    \\[0.15cm]
       $p_1=10$ &Cover & 0.92(0.03) & 0.87(0.02)& 0.99(0.00) & \\
       &$\sqrt{n}$Length &  5.14(0.03) & 3.02(0.04)& 6.80(0.06) &  \\
         & $\sqrt{n}$Bias &-1.04(0.89) & 1.41(1.22) &--- &\\
     \hline\hline
     \end{tabular}
     }
     \end{adjustbox}
      \begin{tablenotes}\small
   \item Note: ``Cover" is the empirical coverage of the 95\% lower bound for $\beta_{\hat{s}}$.  `` $\sqrt{n}$Bias " captures the root-$n$ scaled Monte Carlo bias for estimating $\beta_{\hat{s}}$, and  `` $\sqrt{n}$Length " denotes the  root-$n$ scaled length of the 95\% lower bound for $\beta_{\hat{s}}$. 
     \end{tablenotes}
 \end{table}

\section{Casual effect identification}

\subsection{Casual effect identification under the proposed model}

In this section, our goal is to showcase that the parameter of interest $\beta$ indeed represents subgroup treatment effects under Model \eqref{supp-eqn:setup}

\begin{equation}\label{supp-eqn:setup}
    \text{logit}\left\{{\mathbb{P}}(\y=1\mid \z,\x)\right\} = \mbox{z}\transpose\beta + \x\transpose\gamma,\quad  \left\|\gamma\right\|_0 \ll  p.
\end{equation}

We work under the Neyman-Rubin \citep{neyman1923application,rubin1974estimating} causal model.  In accordance with our case study design, each subject is either randomly assigned the treatment, meaning that nature has assigned at least one copy of rs12916-T allele, or the control, meaning that the subject does not inherit rs12916-T allele. The potential outcome $\y(1)$ ($ \y(0)$) is the potential T2D status we would have observed if the subject carries (does not carry) rs12916-T allele. The observed outcome $\y = \mathbf{1}(\text{the subject}\text{ is diagnosed with T2D})$ thus equals $\y = \t\y(1) + (1-\t)\y(0)$. We work under the stable unit treatment value assumption (SUTVA) and the unconfoundedness assumption listed below.

\begin{assumption}{10}{}
 If unit $i$ receives treatment $\t_i$, the observed outcome $\y_i$ equals the potential outcome $\y_i(\t_i)$. In other words, the potential outcome for unit $i$ under treatment $\t_i$ is unrelated to the treatment received by other units.
\end{assumption}

\begin{assumption}{11}{}
Conditional on a set of potential confounders $\w$, the treatment is independent with the potential outcomes, that is $\t\independent\y(1),\y(0) | \w$.
\end{assumption}

Since we are interested in the subgroup treatment effect, we use $\s$ to denote subgroup indicator variables. We consider six non-overlapping subgroups, $\s\in\{1,2,3,4,5,6\}$.   Because the potential outcomes are binary random variables, we quantify our causal parameter of interest in each subgroup using log odd ratios. For the heterogeneous treatment effect in subgroup $s\in\{1,\ldots,6\}$, the causal parameter of interest is defined as 
\begin{align*}
\log \beta_{s} &= \log \frac{ \mathbb{P}\big( \y(1)=1|\s=s\big)/ \big[1-\mathbb{P}\big( \y(1)=1|\s=s\big)\big]}{ \mathbb{P}(\y(0) = 1|\s=s)/\big[1-\mathbb{P}\big( \y(0)=1|\s=s\big)\big]},\\
&=: \text{logit}\{ \mathbb{P}\big(\y(1)=1|\s\big)\} - \text{logit}\{ \mathbb{P}\big(\y(0)=1|\s\big)\},
\end{align*}

The key challenge in causal inference is that for each subject we only observe their potential outcomes under one of the two possible treatments, but never both. Since the potential outcomes are not observed as a priori, our study design aims to enhances the plausibility of the ``unconfoundedness assumption" so that causal effects can be identified. The so called unconfoundedness assumption ensures that conditional on a set of potential confounders $\w$, the treatment is independent with the potential outcomes, that is $\t\independent\y(1),\y(0) | \w$. 

Our study design enhances the plausibility of the unconfoundedness assumption from two perspectives. On the one hand, the treatment $\t$ is a genetic variant, which is randomly inherited at conception and is not associated with Type 2 diabetes according to GWAS Catalog, therefore it might be reasonable to expect the treatment variable is independent of the potential T2D status acquired after birth (i.e., $\t\independent\y(1),\y(0)$). On the other hand, if one believes that the causal effect between the treatment and the outcome might still be confounded, including prior-birth features (such as age, race, and genetic variant information) as potential confounders makes unconfoundedness assumptions more plausible. 

Under the unconfoundedness assumption, we can then identify the causal parameter of interest by conditioning on the confounders $\w$. Take the conditional potential risk in the treated group for subgroup $s$ for example, we identify this causal parameter with 
\begin{align*}
    \mathbb{P}\big( \y(1)=1|\s=s\big) = \mathbb{E}_{\w}\big[ \mathbb{P}(\y=1|\t = 1, \s=s, \w) \big], \quad  \ s\in\{1,
    \ldots,6\}. 
\end{align*}

Given the identification condition above, the conditional mean of the outcome model is unknown and therefore needs to be modeled and estimated. In the presence of many potential confounders, we assume that the conditional potential outcome model satisfies
\begin{align}\label{eq:potential-outcome-model}
    \text{logit}\big\{ \mathbb{P}\big(\y(t)=1|\w, \s \big)\big\} &=  \text{logit}\big\{\mathbb{P}(\y=1|\t = t, \s=s, \w)\big\},\\
      &=\delta_0+t\delta_1\transpose s +
     \delta_2\transpose s + \delta_3\transpose \w,
\end{align}
where $\w$ includes age, race, and genetic variants associated with T2D related factors (including LDL, high density lipoprotein and obesity). Model \eqref{eq:potential-outcome-model} captures our prior belief that the treatment effects can be heterogeneous across the pre-specified subgroups, but may not differ across subpopulations with different genotypes. Model \eqref{eq:potential-outcome-model} is equivalent to the following logistic regression model with interactions
\begin{align*}
     \text{logit}\{\mathbb{P}(\y=1|\w,\s, \t)\}
     &=: \z\transpose\beta + \x\transpose\gamma =: \text{logit}\{\mathbb{P}(\y=1|\z,\x)\},
\end{align*}
where $\beta = \big(\log \alpha_1, \ldots, \log \alpha_6\big)$ indeed represents  the subgroup treatment effects. $\z\transpose = \t\s\transpose$, where $ \s = \big(\mathds{1}(\s_i=1),\ldots,\mathds{1}(\s_i=6)\big)$. $\x\transpose = (\mathbf{1}\transpose, \tilde{\s}\transpose, \w\transpose)$, $ \tilde{\s} = \big(\mathds{1}(\s_i=1),\ldots,\mathds{1}(\s_i=5)\big)$, $\gamma = (\delta_0, \delta_2\transpose, \delta_3\transpose)\transpose$. The derivations are provided in Section \ref{subsec:identification-proof-two} for two-subgroup case and in \ref{subsec:identification-proof-six} for six-subgroup case.

\subsection{Parameter identification proof: two subgroups}\label{subsec:identification-proof-two}

\begin{proof}
Assume  an i.i.d. random sample $\{\y_i,\t_i,\s_i,\w_i\}_{i=1}^n$, $\s_i \in\{0,1\}$. Assume  $\text{logit}(\mathbb{P}[\y_i=1|\w_i,\s_i, \t_i]) =\delta_0 + \delta_1\t_i\s_i+ \delta_2\t_i(1-\s_i) + \delta_3\s_i + \delta_4\transpose \w_i$. 
\begin{align*}
   \text{logit}\Big( \mathbb{P}[\y_i(1)=1|\w_i, \s_i]\Big) &= \delta_0 + \delta_1\s_i+ \delta_2(1-\s_i) + \delta_3\s_i + \delta_4\transpose \w_i ,\\
   \text{logit}\Big( \mathbb{P}[\y_i(0)=1|\w_i, \s_i]\Big) &= \delta_0+\delta_3\s_i + \delta_4\transpose \w_i,\\
   \log \alpha_{\s}  &=  \log \Big(\frac{ \mathbb{P}\big( \y(1)=1|\s_i\big)/ \big[1-\mathbb{P}\big( \y(1)=1|\s_i\big)\big]}{ \mathbb{P}(\y(0) = 1|\s_i)/\big[1-\mathbb{P}\big( \y(0)=1|\s_i\big)\big]}\Big),\\
   &= \text{logit}\Big( \mathbb{P}[\y_i(1)=1| \s_i]\Big) - \text{logit}\Big( \mathbb{P}[\y_i(0)=1|\s_i]\Big),\\
   &= \delta_1\s_i + \delta_2(1-\s_i),\\
  \text{logit}(\mathbb{P}[\y_i=1|\w_i,\s_i, \t_i]) &=\t_i\text{logit}\Big( \mathbb{P}[\y_i(1)=1|\w_i, \s_i]\Big) + (1-\t_i) \text{logit}\Big( \mathbb{P}[\y_i(0)=1|\w_i, \s_i]\Big),\\
  &=\t_i(\delta_0 + \delta_1\s_i+ \delta_2(1-\s_i) + \delta_3\s_i + \delta_4\transpose \w_i) + (1-\t_i)(\delta_0+\delta_3\s_i + \delta_4\transpose \w_i) ,\\
  &= \t_i\delta_0 + \t_i\delta_1\s_i + \t_i\delta_2(1-\s_i) + \t_i\delta_3\s_i + \t_i\delta_4\transpose\w_i\\ 
  &+(1-\t_i)\delta_0 + (1-\t_i)\delta_3\s_i + (1-\t_i)\delta_4\transpose\w_i,\\
  &= \delta_0 + \t_i\s_i\log\alpha_1 + \t_i(1-\s_i)\log\alpha_0 + \delta_3\s_i + \delta_4\transpose\w_i,\\
\text{logit}(\mathbb{P}[\y_i=1|\w_i,\s_i,\t_i]) &= \underbrace{\begin{pmatrix}
\t_i\s_i & \t_i (1-\s_i)
\end{pmatrix}}_{\z_i\transpose}
\underbrace{\begin{pmatrix}
\log \alpha_1\\
\log \alpha_0
\end{pmatrix}}_{\beta}+ 
\underbrace{\begin{pmatrix}
1 &  \s_i & \w_i\transpose
\end{pmatrix}}_{\x_i\transpose}
\underbrace{\begin{pmatrix}
\delta_0\\
\delta_3\\
\delta_4
\end{pmatrix}}_{\gamma}.
\end{align*}

The above model is thus equivalent to 
\begin{align*}
  \text{logit}\{\mathbb{P}[\y_i=1|\z_i,\x_i]\}  &= \z_i\transpose\beta + \x_i\transpose\gamma,
\end{align*}
where $\z_i$ contains the subgroup-treatment interaction terms. $\beta = (\log\alpha_1,\ \log \alpha_0)\transpose$ represents the subgroup treatment effects. $\x_i$ contains an intercept, potential confounders, and subgroup indicator variables. Therefore, the subgroup parameter of interest, $\beta$, is identifiable under our proposed model.
\end{proof}

\subsection{Parameter identification proof: six subgroups}\label{subsec:identification-proof-six}

\begin{proof}
Define $\s_i$ as a vector of subgroup indicators, $\s_i\in\mathbb{R}^6$. Assume  $\text{logit}(\mathbb{P}[\y_i=1|\w_i,\s_i, \t_i]) =\delta_0 +\t_i\delta_1\transpose\s_i + \delta_2\transpose\s_i + \delta_3\transpose \w_i$.
\begin{align*}
   \text{logit}\Big( \mathbb{P}[\y_i(1)=1|\w_i, \s_i]\Big) &= \delta_0 + \delta_1\transpose\s_i+ \delta_2\transpose\s_i + \delta_3\transpose \w_i ,\\
   \text{logit}\Big( \mathbb{P}[\y_i(0)=1|\w_i, \s_i]\Big) &= \delta_0+\delta_2\transpose\s_i + \delta_3\transpose \w_i,\\
   \log \alpha_{\s}  &=  \log \Big(\frac{ \mathbb{P}\big( \y(1)=1|\s_i\big)/ \big[1-\mathbb{P}\big( \y(1)=1|\s_i\big)\big]}{ \mathbb{P}(\y(0) = 1|\s_i)/\big[1-\mathbb{P}\big( \y(0)=1|\s_i\big)\big]}\Big),\\
   &= \text{logit}\Big( \mathbb{P}[\y_i(1)=1| \s_i]\Big) - \text{logit}\Big( \mathbb{P}[\y_i(0)=1|\s_i]\Big),\\
   &= \delta_1\transpose\s_i,\\
  \text{logit}(\mathbb{P}[\y_i=1|\w_i,\s_i, \t_i]) &=\t_i\text{logit}\Big( \mathbb{P}[\y_i(1)=1|\w_i, \s_i]\Big) + (1-\t_i) \text{logit}\Big( \mathbb{P}[\y_i(0)=1|\w_i, \s_i]\Big),\\
  &=\t_i(\delta_0 + \delta_1\transpose\s_i  + \delta_2\transpose\s_i + \delta_3\transpose \w_i) + (1-\t_i)(\delta_0+\delta_2\transpose\s_i + \delta_3\transpose \w_i) ,\\
  &= \t_i\delta_0 + \t_i\delta_1\transpose\s_i  + \t_i\delta_2\transpose\s_i + \t_i\delta_3\transpose\w_i\\ 
  &+(1-\t_i)\delta_0 + (1-\t_i)\delta_2\transpose\s_i + (1-\t_i)\delta_3\transpose\w_i,\\
  &= \delta_0 + \t_i\s_i\transpose\log\alpha_{s_i} +  \delta_2\transpose\s_i + \delta_3\transpose\w_i,\\
  \text{Let} \ \s_i &= \big(\mathds{1}(\s_i=1),\ldots,\mathds{1}(\s_i=6)\big)\transpose, \ \tilde{\s}_i =\big(\mathds{1}(\s_i=1),\ldots,\mathds{1}(\s_i=5)\big)\transpose\\ \log\alpha_{s_i} &= \big(\log\alpha_1,\ldots,\log\alpha_6\big)\transpose,\\
\text{logit}(\mathbb{P}[\y_i=1|\w_i,\s_i,\t_i]) &= \underbrace{
(\t_i\s_i\transpose)
}_{\z_i\transpose}
\underbrace{
(\log \alpha_{s_i})}_{\beta}+ 
\underbrace{
(1, \tilde{\s}_i\transpose, \w_i\transpose)}_{\x_i\transpose}
\underbrace{
(\delta_0,\delta_3,\delta_4)
}_{\gamma}.
\end{align*}
The above model is thus equivalent to Model (1) considered in the main manuscript 
\begin{align*}
  \text{logit}\{\mathbb{P}[\y_i=1|\z_i,\x_i]\}  &= \z_i\transpose\beta + \x_i\transpose\gamma,
\end{align*}
where $\z_i$ contains the subgroup-treatment interaction terms. $\beta = (\log\alpha_1,\ldots, \log \alpha_6)$ contains the subgroup parameter of interest. $\x_i$ contains an intercept, subgroup indicators, and potential confounders. Therefore, the parameter of interest $\beta$ represents subgroup treatment effect under the proposed model.
\end{proof}

\section{Additional real data results}

\subsection{One-sided lower bounds}

In this section, we show the real data results with one-sided confidence lower bound. From Table \ref{table:real-data-one-sided}, the results of the R-Split estimator without bootstrap calibration suggest that the high-genetic-risk female subgroup is the most vulnerable group for developing T2D with estimated log-odds ratio $0.41$, with $p$-value $0.030$, and 95\% one-sided confidence lower bound $0.10$ (${\rm OR}=1.11$) with $p$-value $0.015$. Our proposed bootstrap assisted R-Split results suggest that among high-genetic-risk female patients, the odds of developing T2D after taking statins are 1.42 times the odds of developing T2D for the patients without taking statins ($p$-value $0.019$ for one-sided test).

\begin{table}[h!]
    \centering
 \begin{adjustbox}{width={\textwidth},center}%
 \begin{tabular}{ccccc}
    \hline
     \hline\\[-2ex] 
       Method & Subgroup (prevalence; \# of case) & Est (95\% LB)  & $p$-value & Bonf $p$-value  \\
  \\[-2ex] 
    \hline
      \\[-2ex] 
   R-Split & High-risk female $(0.14,100)$  & $0.41~(0.10)$ & $0.015$  & $0.090$  \\
    \\[-2ex] 
  (without bootstrap calibration)&Mid-risk female $(0.12,396)$ &  $0.10~(-0.01)$ & $0.066$ & $0.396$ \\
  \\[-2ex] 
  &Low-risk female $(0.11,630)$ & $-0.00~(-0.08)$ & $0.527$ & $1$ \\
  \\[-2ex]
  & High-risk male $(0.24,139)$  & $-0.07~(-0.33)$ & $0.664$  & $1$  \\
    \\[-2ex] 
  &Mid-risk male $(0.21,561)$ & $0.02~(-0.06)$ & $0.336$ & $1$ \\
  \\[-2ex] 
  &Low-risk male $(0.17,739)$  &  $-0.03~(-0.14)$ & $0.667$ & $1$\\
  \\[-2ex] 
    &Overall  & $0.07~(-0.12)$ &  $0.267$ & --\\
    \\[-2ex] 
        
         \\[-2ex] 
         Simultaneous & High-risk female $(0.14,100)$ & -- &  $0.127$ & --\\
          \\[-2ex] 
         \hline
         \\[-2ex] 
    Bootstrap-assisted R-Split & High-risk female $(0.14,100)$ & $0.35~(0.07)$ &  $0.019$ & --\\
          \\[-2ex] 
     \hline\hline
    \end{tabular}
    \end{adjustbox}
    \caption{\label{table:real-data-one-sided} Estimated treatment effects (Est) on the PHS cohort in six subgroups divided by gender and T2D genetic risk, together with the $95\%$ confidence lower bound (LB), the corresponding $p$-values and the Bonferroni $p$-values in the last column. We also present the prevalence of T2D in each subgroup.}
\end{table}

\subsection{Calibration of the second most vulnerable subgroup}

 As a secondary analysis, we first remove the high-risk female subgroup, which is the most vulnerable subgroup out of the six non-overlapping subgroups. Then the second most vulnerable subgroup (mid-risk female) now becomes the most vulnerable subgroup among the remaining five subgroups. We apply our method to calibrate the estimated treatment effect of the mid-risk female subgroup. The results are summarized in Table \ref{tab:second-largest}. Table \ref{tab:second-largest} suggests that our method can be naturally applied to correct for the winner's curse bias on the second largest coefficient. After calibration, the mid-risk female subgroup remains to be non-significant. 

\begin{table}[htb!]
    \centering
\begin{adjustbox}{width={\textwidth},center}%
 \begin{tabular}{ccccc}
    \hline
     \hline\\[-2ex] 
       Method & Subgroup (prevalence; \# of case) & Est (95\% CI)  & $p$-value & Bonf $p$-value  \\
  \\[-2ex] 
    \hline
      \\[-2ex] 
   R-Split&Mid-risk female $(0.12,396)$ &  $0.11~(-0.02,~0.23)$ & $0.10$ & $0.52$ \\
  \\[-2ex] 
 (without bootstrap calibration)  &Low-risk female $(0.11,630)$ & $-0.00~(-0.16,~0.15)$ & $0.96$ & $1$ \\
  \\[-2ex]
  & High-risk male $(0.24,139)$   & $-0.08~(-0.42,~0.26)$ & $0.64$ & $1$  \\
    \\[-2ex] 
  &Mid-risk male $(0.21,561)$  & $0.02~(-0.08,~0.13)$ & $0.67$ & $1$ \\
  \\[-2ex] 
  &Low-risk male $(0.17,739)$   &  $-0.04~(-0.14,~0.06)$ & $0.50$ & $1$ \\
  \\[-2ex] 
         \hline
         \\[-2ex] 
    Bootstrap-assisted R-Split & Mid-risk female $(0.12,396)$ & $0.04~(-0.18,~0.27)$ & $0.36$  & --\\
          \\[-2ex] 
     \hline\hline
    \end{tabular}
    \end{adjustbox}
    \caption{\label{tab:second-largest} Estimated treatment effects (Est), their two-sided $95\%$ CI, corresponding two-sided $p$-values and the Bonferroni $p$-values after removing the high-risk female subgroup.}
\end{table}

\section{Comparison of pre-defined and post-hoc identified subgroups}

In our main manuscript, ``pre-defined subgroups" refers to candidate subgroups that are defined based on prior knowledge, while ``post-hoc identified subgroups" refers to the subgroups identified via data-adaptive identification procedures. Typically, pre-defined subgroups bear better 
interpretability than post-hoc identified subgroups and avoid the potential bias issue induced by data-adaptively identifying candidate subgroups. Therefore, one considers post-hoc identified subgroups when there is no prior knowledge on subgroup segregation. In our case study, we consider pre-defined subgroups because previous studies \citep{mora2010statins,waters2013cardiovascular} suggest that T2D risk might be heterogeneous across sex and T2D genetic profiles.

Although the SNPs we use to define the subgroups are pre-specified using prior knowledge independent of the data, we are able to identify some individualized treatment effects (ITE) and the corresponding subgroups that are similar to our pre-specified subgroups by applying some existing methods to our data. As an illustrative example, we perform logistic regression with Lasso penalty for the model
\begin{equation}\label{eq:model-data-adaptive}
    \mbox{T2D}\sim \mbox{Age}+\mbox{Ethnicity}+\mbox{Treatment}+\mbox{SNPs}+\mbox{Treatment * SNPs},
\end{equation}
on all the females. The above model has been widely adopted in subgroup identification literature \citep{imai2013estimating,dusseldorp2014qualitative}. We apply Model \eqref{eq:model-data-adaptive} on females because we aim to investigate if the data-adaptively identified female subgroup also exhibits significant heterogeneous treatment effect as the pre-defined female subgroup in the main manuscript. Recall that in the main manuscript, we observe  significant treatment effect only in the female high-T2D-risk subgroup. ``High-T2D-risk" was defined by the number of risk alleles of rs35011184-A and rs1800961-T. 

Here, ``SNPs" represents the indicators of having $1$ or $2$ risk alleles of all the 329 SNPs considered in our case study.
Denote the fitted coefficients for the predictors in $\{\mbox{Treatment * SNPs}\}$ as $\widehat\zeta$. We use $\widehat\zeta$ to combine and calculate a linear score of SNPs for each individual as the ``individualized treatment effect" (ITE), which characterizes the treatment effect heterogeneity of statin usage across different genetic profiles. 
A higher ITE score represents a higher genetic risk of developing T2D when treated with statins. From the data, we are able to identify a ``high genetic risk" subgroup as those subjects with the top $12\%$ (i.e. prevalence of T2D in the female population) ITE scores based on Model \eqref{eq:model-data-adaptive}. The data-adaptively identified subgroup carries the similar clinical implication to our pre-defined subgroup, that is the female patients who have higher baseline genetic risks of developing T2D.

When comparing the data-adaptively identified subgroup with our pre-defined female high-risk subgroup (i.e. the females with $\geq 2$ T2D risk alleles), interestingly, we find that the odds ratio of this data-adaptively identified high-risk group against our pre-defined high-risk subgroup is as high as $2.8$, ($95\%$ CI: $(2.3,\  3.4)$, $p$-value $<10^{-16}$).
The results demonstrate that the  pre-defined subgroup and the data-adaptively identified subgroup are similar. 

Please note that identifying subgroups data-adaptively might bring another source of bias. In our main manuscript, we adopt the subgroups defined based on prior knowledge for analyses which can help avoid such post-selection bias issues. Because correcting the selection bias in data-adaptively identified subgroups is not the main objective of our manuscript,  we shall leave the development of data-adaptive methodologies to future research.

\section{Discussions on other possible causal pathways}

Kindly pointed out by an anonymous referee, there might be patients who did not carry the genetic variant rs12916-T but did take statins to treat diseases such as CAD. In what follows, we shall demonstrate that although there exist patients who did not carry this variant but did take statins, our current study design still provides valid causal effect estimates. 

\begin{figure}  
    \centering
 \includegraphics[width=0.5\textwidth]{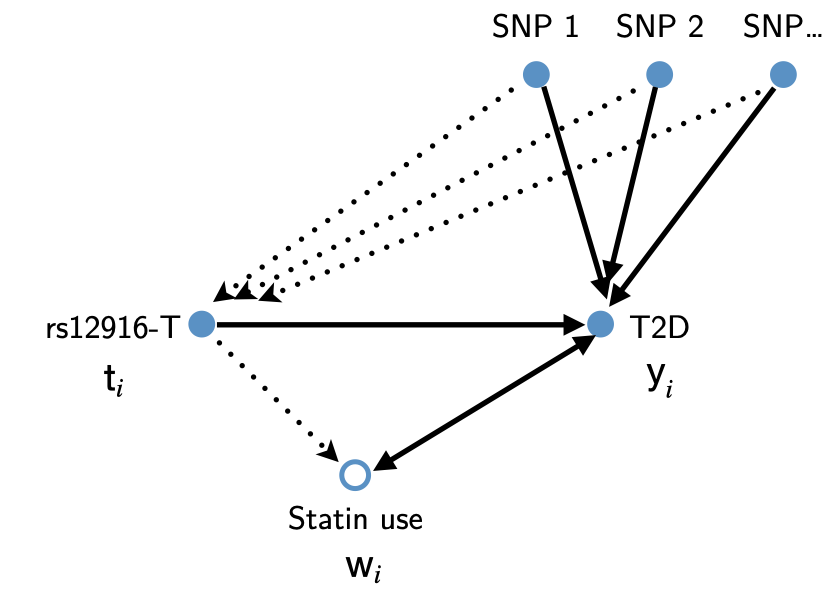}\caption{ The causal diagram of incorporating statin use information. ``Statin use" is a mediator in this causal diagram  }
    \label{fig:causal-diagram}
\end{figure}

Given that our study cohort may contain patients who did not carry the variant rs12916-T but did take statins, including these patients in our causal analysis potentially opens two different causal pathways including $\t_i\rightarrow\w_i\leftarrow\y_i$ and $\t_i\rightarrow\w_i\rightarrow\y_i$, where $\w_i$ represents statin use information after birth. We reflect these additional pathways in the causal diagram in Figure \ref{fig:causal-diagram}. In the causal diagram provided in Figure \ref{fig:causal-diagram}, there are three causal pathways started from $\t_i$ and ended in $\y_i$, including $\t_i\rightarrow\w_i\rightarrow\y_i$, $\t_i\rightarrow\w_i\leftarrow\y_i$, and $\t_i\leftarrow \text{SNP}_j\rightarrow\y_i$, where $\w_i$ represents statin use information after birth.
In what follows, we shall analyze each pathway and demonstrate that only our current causal pathway $\t_i\leftarrow \text{SNP}_j\rightarrow\y_i$ in this case study provide valid causal conclusion. 

By the D-separation criteria (\cite{glymour2016causal}, Ch.3), there are two creditable causal pathways ($\t_i\rightarrow\w_i\rightarrow\y_i$ and $\t_i\leftarrow \text{SNP}_j\rightarrow\y_i$) present in Figure \ref{fig:causal-diagram}, meaning that conducting causal analysis with either of them would lead to valid causal conclusions. However, only the  pathway $\t_i\leftarrow \text{SNP}_j\rightarrow\y_i$ allows us to establish valid causal conclusion in our study cohort.

For the pathway $\t_i\rightarrow\w_i\rightarrow\y_i$, under appropriate conditions provided in 
\cite{imai2010identification}, this pathway indeed allows us to estimate the direct effect of inheriting rs12916-t on T2D risk. Unfortunately, because statin use information collected after birth is not available in our EHR data, we were not able to conduct causal inference following this pathway. 

For the pathway $\t_i\leftarrow \text{SNP}_j\rightarrow\y_i$, not only all confounder information is available to us in the EHR data, 
our study design also guarantees that the cause (carrying rs12916-T [proxy for pharmacological action of statin use] or not) must occur before T2D onset as genetic variants are randomly inherited at conception. Therefore, we are able to establish a clear causal direction between the treatment and the outcome. On top of valid causal directions, following the causal parameter identification proof in Supplementary Materials Section E, we guarantee that we are able to estimate the causal effect from this causal pathway. 

For the pathway $\t_i\rightarrow \w_i\leftarrow\y_i$, because $\w_i$ is a collider, including this variable in the causal analysis will create a non-causal association between $\t_i$ and $\y_i$. This pathway thus should not be considered in our analysis. 

Furthermore, we want to note that even if statin use information is available, incorporating statin use information under our electronic health records (EHR) data is infeasible. This is because the temporal precedence between statin use and T2D onset is not available, the causal direction between $\t_i$ and $\w_i$ can not be established. In sum, we are not able to decide if statin use is a collider or a mediator between $\t_i$ and $\y_i$.

Lastly, while rs12916-T is sometimes used as an instrumental variable in Mendelian Randomization analyses \citep{wurtz2016metabolomic}, we use rs12916-T as a surrogate treatment variable, instead of an instrumental variable. Our study design is thus different from Mendelian randomization (MR) \citep{kang2016instrumental,windmeijer2019use}. Furthermore, because MR often assumes that the causal direction between two traits is known as a priori \citep{xue2020inferring} and our data neither contain statin use information after birth nor provide the temporal order between statin usage and T2D onset, the causal direction for MR analyses can not be specified and MR is not suitable in our case study.

\section{Sensitivity analysis on the surrogate outcome}\label{subsec:sensitivity-outcome}
Given that our outcome is an error-prone surrogate of the true disease status,  we conduct a sensitivity analysis regarding the potential misspecification of the logistic regression model for the true EHR disease status against the covariates. This sensitivity analysis is inspired by \cite{hong2019semi} and \cite{zhang2020maximum}. Denote $\y_i$ as the observed EHR surrogate of T2D disease status, $\y^*_i$ as the unobserved true disease status of T2D.
Following \cite{hong2019semi}, we assume that
\begin{equation}\label{eq:surrogate-logistic-model}
    \text{logit}\left\{{\mathbb{P}}(\y^*=1\mid \z,\x)\right\} = \mbox{z}\transpose\beta + \x\transpose\gamma,\quad \y\perp (\z,\x)\mid \y^*.
\end{equation}
Eq \eqref{eq:surrogate-logistic-model} entails two model assumptions. First, the true disease status $\y^*$ follows a logistic regression model against the subgroup-treatment interaction terms $\z$ and baseline covariates $\x$ (e.g. genetic and demographic variables). Second, by conditioning on $\y^*$, the surrogate outcome $\y$ is independent of the predictors, that is the surrogate outcome obtained from phenotyping algorithms is only related to the baseline covariates $\x$ through $\y^*$.  Under these model assumptions, the log-likelihood function for $\{ (\y_i, \x_i, \z_i) \}_{i=1}^n$ can be written as
\begin{equation}
\begin{split}
\mathcal{L}\left(  \beta,\gamma,\mu\right)  =\frac{1}{n}\sum_{i=1}^{n}&\y_i\log\left\{
\mu_1g(\mbox{z}_i\transpose\beta + \x_i\transpose\gamma)+\mu_0\bar g(\mbox{z}_i\transpose\beta + \x_i\transpose\gamma)\right\}\\
&+(1-\y_i)\log\left\{
(1-\mu_1)g(\mbox{z}_i\transpose\beta + \x_i\transpose\gamma)+(1-\mu_0)\bar g(\mbox{z}_i\transpose\beta + \x_i\transpose\gamma)\right\},    
\end{split}
\label{equ:llh}
\end{equation}
where $\mu=(\mu_0,\mu_1)$, $\mu_0={\mathbb{P}}(\y=1\mid\y^*=0)$, $\mu_1={\mathbb{P}}(\y=1\mid\y^*=1)$, $g(\cdot)=\text{logit}^{-1}(\cdot)$, and $\bar g(\cdot)=1-g(\cdot)$. Due to the high dimensionality of $\gamma$, we introduce Lasso penalty and adopte an EM algorithm to solve
\[
\{\widehat\beta,\widehat\gamma,\widehat\mu\}=\arg\min_{\left(  \theta,\beta\right)
}\left\{  -\mathcal{L}\left(  \beta,\gamma,\mu\right)  +\lambda\Vert\gamma\Vert
_{1}\right\},
\]
where $\lambda$ is the penalty parameter. Details of the EM algorithm can be found in \cite{hong2019semi}. Then we can derive the conditional mean of the true T2D status $\y_i^*$ given $(\y_i, \x_i, \z_i)$ by
\[
\hat{\y}^*_i=\frac{\mu_1^{\y_i}(1-\mu_1)^{1-\y_i}g(\mbox{z}_i\transpose\widehat\beta + \x_i\transpose\widehat\gamma)}{\mu_1^{\y_i}(1-\mu_1)^{1-\y_i}g(\mbox{z}_i\transpose\widehat\beta + \x_i\transpose\widehat\gamma)+\mu_0^{\y_i}(1-\mu_0)^{1-\y_i}\bar g(\mbox{z}_i\transpose\widehat\beta + \x_i\transpose\widehat\gamma)}.
\]
Finally, for calibration of the error-prone surrogate outcome $\y_i$, we sample the T2D status $\widetilde \y^*_i$ following ${\mathbb{P}}(\widetilde \y^*_i=1\mid \y_i,\z_i,\x_i)=\widehat{\y}^*_i$, for $i=1,2,\ldots,n$, and implement the bootstrap-assisted R-split with $\widetilde \y^*_i$ against $(\z_i,\x_i)$. Because this sensitivity analysis uses the calibrated $\widetilde \y^*_i$ instead of $\y_i$ as the outcome, the sensitivity analysis can correct for the approximation error of the EHR outcome $\y_i$ to the true disease status $\y_i^*$ \citep{hong2019semi}.

To avoid over-fitting bias induced by estimating parameters ($\{\widehat\beta,\widehat\gamma,\widehat\mu\}$) and constructing $\widehat{\y}^*_i$ on the same data, we use a cross-fitting strategy that splits the data into five folds, estimates $\{\widehat\beta,\widehat\gamma,\widehat\mu\}$ leaving out one fold each time, and constructs each $\widehat{\y}^*_i$ on the left-out fold with the independent estimators. We replicate the sampling of $\{\widetilde \y^*_i:i=1,2,\ldots,n\}$ for $10$ times 
and take the average over estimated $\beta_{\max}$'s and the associated standard errors under $(\widetilde{\y}^*_i,\z_i,\x_i)$. The resulted point estimates, confidence intervals, and $p$-values are presented in Table \ref{table:sensitivity-outcome}.

\begin{table}[htb!]
    \centering
\begin{adjustbox}{width={\textwidth},center}%
 \begin{tabular}{ccccc}
    \hline
     \hline\\[-2ex] 
       Method & Subgroup (prevalence; \# of case) & Est (95\% CI)  & $p$-value & Bonf $p$-value  \\
  \\[-2ex] 
    \hline
      \\[-2ex] 
   R-Split & High-risk female $(0.14,100)$  & $0.36~(0.05,~0.67)$ & $0.024$  & $0.144$  \\
    \\[-2ex] 
   (without bootstrap calibration) &Mid-risk female $(0.12,396)$ &  $0.09~(-0.07,~0.25)$ & $0.275$ & $1$ \\
  \\[-2ex] 
  &Low-risk female $(0.11,630)$ & $0.03~(-0.09,~0.15)$ & $0.651$ & $1$ \\
  \\[-2ex]
  & High-risk male $(0.24,139)$  & $-0.05~(-0.36,~0.26)$ & $0.754$  & $1$  \\
    \\[-2ex] 
  &Mid-risk male $(0.21,561)$ & $-0.01~(-0.14,~0.13)$ & $0.940$ & $1$ \\
  \\[-2ex] 
  &Low-risk male $(0.17,739)$  &  $-0.01~(-0.13,~0.11)$ & $0.886$ & $1$\\
  \\[-2ex] 
         \hline
         \\[-2ex] 
    Bootstrap-assisted R-Split & High-risk female $(0.14,100)$ & $0.28~(0.02,~0.54)$ &  $0.037$ & --\\
          \\[-2ex] 
     \hline\hline
    \end{tabular}
    \end{adjustbox}
    \caption{\label{table:sensitivity-outcome} The sensitivity analysis of our surrogate outcome, including the estimated treatment effects (Est), their two-sided $95\%$ confidence intervals (CI), the two-sided $p$-values, and the Bonferroni $p$-values obtained by implementing the R-Split and the Bootstrap-assisted R-split procedures with the calibrated outcome $\widetilde \y^*_i$ (instead of $\y_i$) against $(\z_i,\x_i)$. The results are produced by averaging over the results from $10$ repetitions of sampling $\widetilde \y^*_i$.}
\end{table}

Comparing the results in Table \ref{table:sensitivity-outcome} with  Table 4 in the main manuscript, we do not observe any significant differences. In both tables, R-split $p$-values of the high-risk female group are around $0.03$, the $p$-values of the remaining subgroups are non-significant, and the bootstrap-assisted R-Split $p$-values are equal to $0.037$, which lead to the same scientific conclusion as in Table 4 in the main manuscript. The results from the sensitivity analysis suggest that the analyses and findings in Table 4 in the main manuscript are not sensitive to the approximation error of the EHR surrogate $\y_i$ to the true T2D status. This is because our EHR outcome $\y$, derived using MAP, shows a very low approximation error to the true T2D status ($\mbox{AUC}=0.99$, $\mbox{specificity}=0.97$, and $\mbox{sensitivity}=0.92$, as was verified using a small set of gold standard labels).

\end{document}